\documentclass[journal]{IEEEtran}
\usepackage{amsmath,amsfonts}
\usepackage{algorithm}
\usepackage{array}
\usepackage[caption=false,font=normalsize,labelfont=sf,textfont=sf]{subfig}
\usepackage{textcomp}
\usepackage{stfloats}
\usepackage{url}
\usepackage{verbatim}
\usepackage{graphicx}
\usepackage{cite}
\usepackage{algpseudocode}
\usepackage{multirow}

\usepackage{xcolor}
\DeclareMathOperator*{\argmin}{arg\,min}
\usepackage[hidelinks]{hyperref}
\def\NoNumber#1{{\def\alglinenumber##1{}\State #1}\addtocounter{ALG@line}{-1}}
\begin{document}

\title{Cross-Domain Federated Semantic Communication with Global Representation Alignment and Domain-Aware Aggregation}

\author{Loc X. Nguyen, Ji Su Yoon, Huy Q. Le, Yu Qiao, Avi Deb Raha, Eui-Nam Huh~\IEEEmembership{Member,~IEEE}, Walid Saad~\IEEEmembership{Fellow,~IEEE}, Yumin Park, Zhu Han~\IEEEmembership{Fellow,~IEEE}, and Choong Seon Hong~\IEEEmembership{Fellow,~IEEE}
\thanks{Loc X. Nguyen, Ji Su Yoon, Huy Q. Le, Yu Qiao, Avi Deb Raha, Eui-Nam Huh, Yumin Park and Choong Seon Hong are with the Department of Computer Science and Engineering, Kyung Hee University, Yongin-si, Gyeonggi-do 17104, Rep. of Korea, e-mail:\{xuanloc088, yjs9512, quanghuy69, qiaoyu, avi, johnhuh, yumin0906, cshong\}@khu.ac.kr.}
\thanks{Walid Saad is with the Department of Electrical and Computer Engineering, Virginia Tech, Alexandria, VA 22305 USA \{e-mail: walids@vt.edu\}.}
\thanks{Zhu Han is with the Department of Electrical and Computer Engineering at the University of Houston, Houston, TX 77004 USA, and also with the Department of Computer Science and Engineering, Kyung Hee University, Seoul, South Korea, 446-701. email:{\{hanzhu22}@gmail.com\}.}
}


\maketitle

\begin{abstract}
Semantic communication can significantly improve bandwidth utilization in wireless systems by exploiting the meaning behind raw data. However, the advancements achieved through semantic communication are closely dependent on the development of deep learning (DL) models for joint source–channel coding (JSCC) encoder/decoder techniques, which require a large amount of data for training. To address this data-intensive nature of DL models, federated learning (FL) has been proposed to train a model in a distributed manner, where the server broadcasts the DL model to clients in the network for training with their local data. However, the conventional FL approaches suffer from catastrophic degradation when client data are from different domains. In contrast, in this paper, a novel FL framework is proposed to address this domain shift by constructing the global representation, which aligns with the local features of the clients to preserve the semantics of different data domains. In addition, the dominance problem of client domains with a large number of samples is identified and, then, addressed with a domain-aware aggregation approach. This work is the first to consider the domain shift in training the semantic communication system for the image reconstruction task. Finally, simulation results demonstrate that the proposed approach outperforms the model-contrastive FL (MOON) framework by $0.5$ for PSNR values under three domains at an SNR of $1$ dB, and this gap continues to widen as the channel quality improves.
\end{abstract}

\vspace{-5mm}
\section{Introduction}
\IEEEPARstart{T}{he} number of connected devices is growing at an exponential pace, and this trend is expected to continue at a faster pace due to the increasing demand for seamless connectivity among users \cite{10856890,10929033}. This rapid proliferation of new devices, coupled with emerging bandwidth-intensive wireless services, will strain the capacity of next-generation wireless systems. Indeed, under a limited bandwidth, next-generation wireless networks must serve a much higher density of users with stricter quality-of-service requirements, such as extremely low latency for online gaming and resource-intensive communication for augmented reality or digital twin applications\cite{11193869,11192484,qiao2025deepseek,10930485}. To accommodate these new devices and applications, various communication techniques have been proposed, including the integration of sensing and communication \cite{9724245}, mmWave, intelligent networking with the assistance of AI/ML technologies\cite{8755300}. Despite tremendous efforts to improve the communication efficiency of traditional communication, its capacity is limited by Shannon's theory, thus making it difficult to cope with the emerging transition. Consequently, the research community is shifting toward the concept of semantic communication \cite{10554663}, which operates beyond the bit-level paradigm.

Instead of focusing solely on bit reconstruction as done in conventional communication systems, semantic communications aims to successfully interpret the meaning of the message. Semantic communication was introduced by Shannon and Weaver back in the 1950s \cite{shannon1998mathematical}, but it did not achieve promising results due to the technology's limitations. Recently, with the rise of deep learning (DL), semantic communication has gained significant attention due to the feasibility of semantic extraction by DL models. Specifically, researchers have deployed the DL models as the joint source-channel encoder/decoder in the semantic communication system, replacing traditional communication. The DL-based semantic communication has demonstrated outstanding performance in reducing the communication bandwidth, while overcoming the ``cliff effect" problem encountered in traditional systems \cite {DJSCCwithfeedback}. The ``cliff effect" refers to the failure in transmission when the channel quality decreases below a threshold. On the other hand, the semantic communication performance is robust to channel noise and improves as the noise decreases. The improvements are the result of DL models being trained on an extensive amount of data, which is normally collected and gathered at the centralized server. However, data transmission to a centralized server can lead to privacy concerns due to potential leakage of sensitive information. Therefore, federated learning (FL) has been introduced as a distributed learning framework without the need for centralizing the data \cite{mcmahan2017communication}.

In FL, a central server actively and iteratively sends the model to be trained to the devices, i.e., the clients, that possess the data. The clients train the model using their local data and transmit the updated models back to the server, which then aggregates them into a global model to achieve better generalization performance\cite{qiao2025towards}.
Several studies have been proposed to enhance the convergence performance of FL with scheduling policies \cite{10041216}, device-to-device communication links \cite{9562522}, or partial model personalization \cite{pillutla2022federated}, whose efforts aim to alleviate the communication resource requirements and enable its deployment in the real world. 
Existing works have applied the FL framework to train semantic communication models for various tasks, including audio reconstruction \cite{9685654}, image reconstruction \cite{10288204,10531097,wu2025personalized}, classification tasks \cite{10937229,10559618,FLfortaskoriented}.

\begin{figure}[t]
\centering
\includegraphics[width=0.5\textwidth]{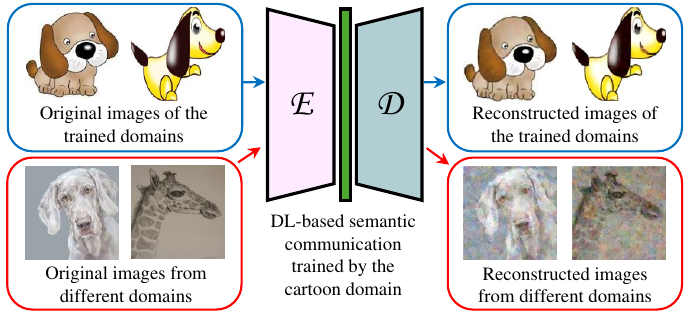}
\caption{The example of domain shift for the DL-based Semantic Communication: the model performs well under the trained domain but fails to obtain good performance for other domains.}
\label{Exampleofcrossdomain}
\end{figure}

However, none of these prior studies considered the cross-domain problem, where client data are from different domains. For example, in the image modality, domains may include photos, art paintings, and human sketches. Each domain has distinct characteristics, and the clients with heterogeneous domain data make the FL process more challenging to converge. This issue may potentially downgrade the system's performance, specifically in our case, the semantic communication model.
In Fig.~\ref{Exampleofcrossdomain}, we provide an example of the domain shift for DL-based joint source-channel coding (D-JSCC) for the reconstruction task. Specifically, we train the semantic model using images from the cartoon domains, and then evaluate the trained model on images from three domains: photo, art, and finally, cartoon. The reconstructed images for the cartoon domain achieve high visual quality, whereas the images from the other domains exhibit colored noise and low performance.

Motivated by this research gap, the main contribution of this paper is the proposal of a novel cross-domain FL framework to train the DL-based joint source-channel encoder/decoder in the semantic communication system. To address the performance degradation of the cross-domain clients in FL\cite{10203389}, we first design a tailored mechanism for the server to achieve a generalization performance by aligning the specialized domain knowledge of clients with global knowledge. Secondly, due to the nature of data domains, some domains have a larger number of data samples, while others are more challenging to obtain, resulting in extremely heterogeneous data samples for clients. This unbalanced data across domains can cause the global model to favor the domain with a large number of samples, thus resulting in poor performance for the remaining domains. Therefore, we tailor an aggregation approach to mitigate this degradation in overall performance. Unlike other studies that consider a classification task \cite{wang2025federated}, our scenario's objective is to reconstruct the original image by training the DL-based semantic communication model. The contributions of our study are provided as follows.
\begin{itemize}
    \item To the best of our knowledge, we are the first to consider the cross-domain clients problem in the FL process for the semantic communication system, which significantly degrades the communication performance in terms of the quality of the reconstructed images. 
    \item We propose a global representation feature to capture the feature generalization of data across the domains. We leverage the global representation feature to align with the local feature in the training process. The proposed alignment approach ensures that each client is aware of the feature discrepancy between its own domain and the global representation, thereby enhancing overall performance.
    \item To address the heterogeneity in the data samples of each domain, we propose a simple yet effective mechanism for global model aggregation, called domain-aware aggregation. Specifically, instead of naively aggregating all client models regardless of their data domains, we first cluster clients within the same domain to formulate a domain-specific model, and then construct a general model across multiple domains.
    \item Finally, we conduct a series of experiments to demonstrate the effectiveness of the aligning mechanism toward the global representation and the effects on performance when conducting the domain-aware aggregation. We validate the results under a variety of FL benchmarks and channel conditions.
\end{itemize}

The rest of the paper is organized as follows. We briefly discuss related works in Section \ref{Related}. Then, Section \ref{System} describes a general system model of the semantic communication and the federated learning framework. Our proposed solution to address the domain shift problem is demonstrated in-depth in Section \ref{Proposed}. We provide the simulation analysis, image visualization in Section \ref{Performance}. Finally, the conclusions are presented in Section \ref{Conclusion}.
\vspace{-2mm}
\section{Related Works}\label{Related}

\subsection{Semantic Communication for Image Reconstruction}
The need for image transmission has never been reduced and continues to increase significantly throughout history, which motivates extensive research on semantic communication for image reconstruction tasks\cite{DJSCCnofeedback,DJSCCAttention,DJSCCSwin,DJSCCDigital}. The authors in \cite{DJSCCnofeedback} were among the first to recognize the potential of the DL model for end-to-end joint source-channel coding, and they proposed two convolutional neural network models for encoder and decoder. Their proposal achieved higher image quality performance when the channel had low signal-to-noise ratio (SNR) values. However, the paper failed to leverage feedback information for the learning process, which motivated the framework in \cite{DJSCCwithfeedback}. Specifically, they exploited the feedback by analyzing the distortion of the transmitted signal, thereby improving the encoding ability of the DL-based model and achieving higher reconstruction quality. On the other hand, the Attention DL-based JSCC (ADJSCC) was proposed to address the dynamics of the channel state information by exposing the DL models to a wide range of SNR values for parameter optimization \cite{DJSCCAttention}. With the development of the architecture for the DL models, the Swin Transformer \cite{liu2021swin} was proposed to resolve the high computing complexity of the attention mechanism, and thus was adopted in \cite{DJSCCSwin} for securing a more efficient system in both computing and communication resources in the semantic communication system. Approaching semantic communication from a different perspective, the authors in \cite{DJSCCDigital} proposed a novel demodulation method to address the dynamic output of the encoder/decoder, thereby bridging the gaps between semantic communication studies and current digital communication systems. However, the aforementioned studies assume the availability of data on a central server for training and overlook the existence of cross-domain data. 

\vspace{-3mm}

\subsection{Cross-domain in Semantic Communication}
There is a limited amount of research on semantic communication for data from different domains. Specifically, the study \cite{10552845} adopted semantic communication for intelligent machine-to-machine applications and proposed a cross-domain interpretation for human inspection. The cross-domain meaning interpretation was obtained by using the Dewey Decimal Classification. In the meantime, a multi-domain adaptive semantic coding network was proposed in \cite{10960416} to effectively encode and decode data from different sources with the assistance of a star-generative adversarial network (Star-GAN) for domain transformations. On the other hand, authors in \cite{10944429} proposed multi-scale semantic communication for the object detection task with the assistance of domain adaptation training, where they consider images captured under different lighting or weather conditions as cross-domain. Unlike these studies, we consider a higher-level cross-domain problem, where the data originates from diverse sources, such as photos, sketches, cartoons, or art images. These cross-domain data are distributed across different semantic communication devices, which presents a challenge for training DL-based models.
\vspace{-2mm}
\begin{figure*}[t]
\centering
\includegraphics[width=0.83\textwidth]{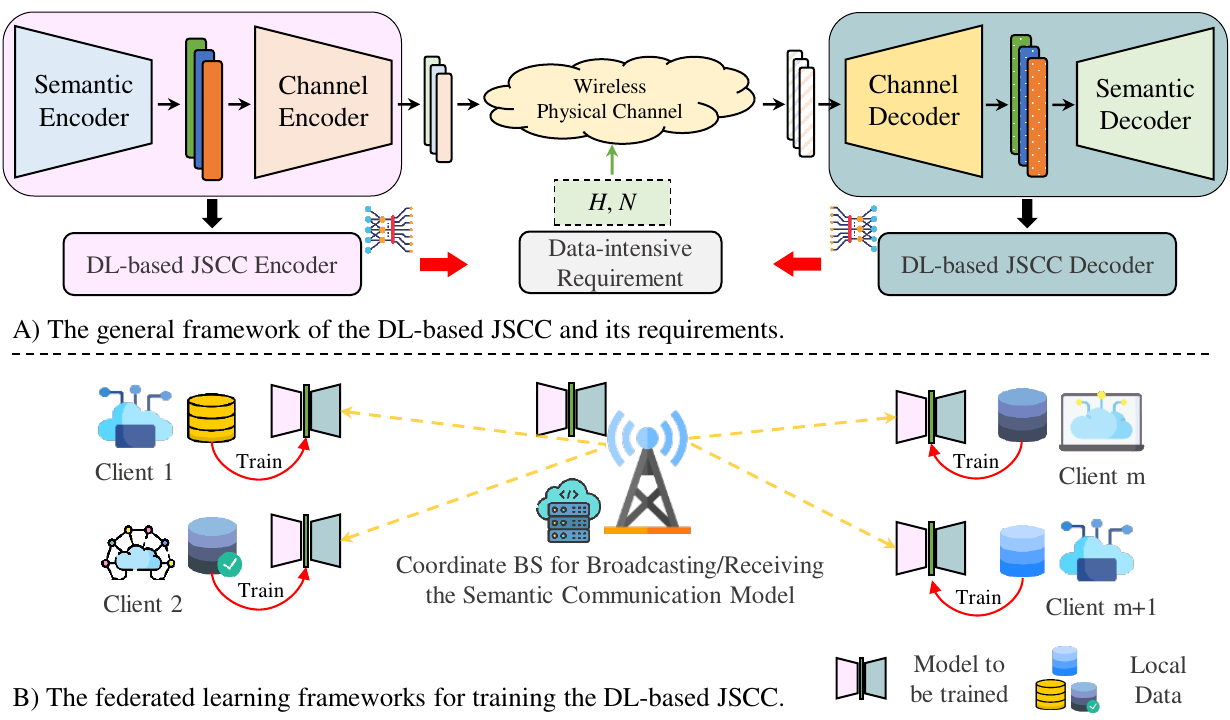}
\caption{A) Provide the general components of the semantic communication system that follow the deep learning-based joint source-channel coding direction. B) Present the implementation of a federated learning framework for training the DL-based semantic communication system, which requires training on a large amount of data.}
\label{SemanticFLmodel}
\end{figure*}
\subsection{Federated Learning for Semantic Communication}
There is an increase in the number of studies on training semantic communication systems with the FL framework, particularly for image transmission \cite{FLSemantic,huh2025feature,FLforGAIassistedSemCom,11096603,zhou2025fedsem}. Specifically, the authors in \cite{FLSemantic} designed a hybrid model combining convolution neural network (CNN) and Transformer designs for DL-based semantic communication and trained it using FL. To actively update the learning model for the system, the authors divided the clients into two groups: one for full parameter updates and one for aligning feature representation, which stabilizes the optimization process \cite{huh2025feature}. Interestingly, the work \cite{FLforGAIassistedSemCom} adopted generative artificial intelligence to assist the FL training of semantic communication by distilling knowledge to the student model, which ensures a unified update direction for the training. In the work \cite{11096603}, the author leveraged cross-modal semantic information to achieve a better user experience, specifically through hybrid coding for the haptic signal. On the other hand, in \cite{zhou2025fedsem}, the authors formulated a problem to jointly optimize the energy consumption and model performance of the federated learning framework by controlling the sub-carrier allocation variable, transmission power, and computing frequency. However, none of the above studies considered the domain drift phenomenon when client data differ across domains, which leads to significant degradation in the performance of the learning system.

\vspace{-3mm}
\section{System Model}  \label{System} 

We consider a set $M$ of devices that can perform a semantic communication process. Before deploying semantic communication, we leverage their local data to train the DL-based joint source-channel coding for image modality. First, we explain the components of the semantic communication model for the transmitter and receiver, and then illustrate the integration of the distributed learning framework to address the data-intensive requirement of our DL-based semantic communication system. Finally, we consider the cross-domain in the client data and discuss its impact on global convergence. 
\vspace{-3mm}
\subsection{Semantic Communication Components and Training}
In Fig.~\ref{SemanticFLmodel}a, we consider the semantic communication model that contains the D-JSCC encoder for the transmitter, a channel model for the wireless environment, and finally a D-JSCC decoder for decoding the signal at the receiver, similar to other semantic studies \cite{DJSCCAttention,DJSCCwithfeedback}. With a given image $I \in$ $\mathbb{R}^{3\times H \times W}$ and a set $\mathcal{S}$ of SNR values, the encoding process is denoted
\begin{equation}
    X_{I}=\mathrm{Enc}_{\alpha}(I,\gamma), \forall I \in \mathbb{R}^{n}, \forall \gamma \in \mathcal{S},
                                                                                                                                                                                                                                                                                                                                                                                                                                                                                                                                                                                                                                                                                                                                                                                                                                                                                                                                                                                                                                                                    \end{equation}
where the $\mathrm{Enc}(\cdot)_{\alpha}$ represents the joint source-channel encoder with its corresponding parameters $\alpha$, $X_{I} \in C^{k}$ is the transmitted signal of the image $I$ through the channel. The compression ratio between the transmitted signal and the original image is the ratio between $k$ and $n$, where $n=$ $3\times H\times W$. The signal is exposed not only to noise but also to fading effects caused by obstacles in the wireless environment:
\begin{equation}
    Y_{I}= H X_{I} + N_{\gamma}, N_{\gamma} \sim  \mathcal{CN}(0,\gamma),
\end{equation}
where $Y_{I}$ and $H$ denote the received signal and the channel fading coefficient at the receiver site, respectively. $N_{\gamma}$ is the channel noise, where its elements follow the complex Gaussian distribution with zero mean and variance $\gamma$. With the received signal $Y_{I}$, the semantic communication receiver decodes it with the D-JSCC decoder as follows:
\begin{equation}
    \hat{I}= \mathrm{Dec}_{\beta}(Y_{I},\gamma), \forall I \in \mathbb{R}^{n}, \forall \gamma \in \mathcal{S}, 
\end{equation}
where $\hat{I}$ and $\mathrm{Dec}_{\beta}(\cdot)$ represents the reconstructed image and the deep joint source-channel decoder with parameters $\beta$ at the receiver, respectively. However, DL-based models are inherently data-hungry, requiring a substantial amount of training data to achieve high performance and robustness against noise. Therefore, the parameters $\theta=\{\alpha,\beta\}$ of our model must be trained with sufficient data before deployment for transmission. Most of the current studies consider a single central server for the end-to-end training as follows:\\
\begin{equation}
        \argmin_{\alpha, \beta}[\mathcal{L}(I,\hat{I})], \forall I \in \mathbb{R}^{n},
\end{equation}
where $\mathcal{L}(\cdot,\cdot)$ is the training loss, determined by the Mean Squared Error (MSE). Centralized training not only consumes a large amount of communication resources due to the massive data transmission, but also raises privacy and sensitivity issues regarding the data collected from clients. Therefore, we conduct the training with the FL framework.
\vspace{-3mm}
\subsection{FL for DL-based Semantic Communication System}
Instead of centralizing the data, each client device actively trains the model with its data, and the BS is responsible for coordinating the local learning process of each device, i.e., the client, by receiving the locally trained model and broadcasting the new aggregated model to clients, as illustrated in Fig.~\ref{SemanticFLmodel}b. Within the study, we focus on image modality; however, we can expand the FL framework to train other modalities such as audio\cite{9685654}, text\cite{10729265}. Each client can possess a different amount of data, which we denote as $D_{m}$, indicating the amount of data that client $m$ has. Upon receiving the global semantic communication model $\theta_{g}$ from the BS server at round $r$, client $m$ trains it with the private images as follows:
\begin{equation} 
\theta_{m}^{r+1} =  \theta_{g}^{r} - \eta \nabla \mathrm{L}_{m}(I,\hat{I}), \forall I \in D_{m}, 
\end{equation}
where $\theta_{m}$ is the updated model of client $m$ after the local training. Then, the BS server conducts the global model aggregation of all the local models with the objective of obtaining better generalization capability, i.e.,
\begin{equation}
    \theta_{g}^{r+1}=\sum_{m=1}^{M} \frac{D_{m}}{D}\cdot\theta_{m}^{r+1}.\label{FedAvg}
\end{equation}

In (\ref{FedAvg}), the contribution of each local client is proportional to the amount of its data relative to the total data of all clients $D$, which is the core of FedAvg \cite{mcmahan2017communication}. We consider the diversity of clients' data, reflected across domains such as real-world photos versus cartoon images. The latent space, characteristics, and underlying distributions of images from different domains are distinct, leading to the creation of derived local models and the degradation of the global model.
\begin{figure*}[t]
\centering
\includegraphics[width=0.95\textwidth]{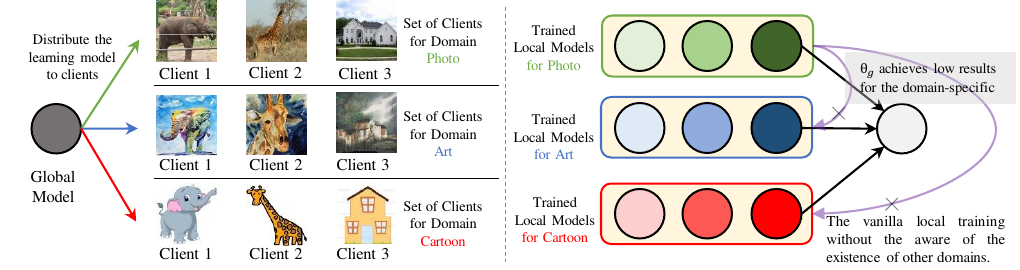}
\caption{Visualization of data from the same class but different domains, showing that their latent features and distributions drift significantly. Each domain can be regarded as a distinct task; therefore, simply aggregating models trained on different tasks may degrade task-specific performance and hinder generalization.}
\label{DomainVisualization}
\end{figure*}
\vspace{-3mm}
\section{Global Representation and Domain-aware Aggregation for Cross-domain FL}\label{Proposed}

We first define $u$ as the source domain of a client, and the total possible set of source domains is $\mathcal{U} =\{u_{1},u_{2},\ldots,u_{n}\}$. Under the conventional training process of the FL framework, each client trained the received global model with its local data without being aware of the existence of other domains' data. Therefore, the clients from different domains will update the parameters accordingly to their own data domain regardless of the training process of the other clients, as shown in Fig.~\ref{DomainVisualization}. This independent updating results in client parameter drift across the network \cite{chen2024fair}. With each client's final local parameters presented by $\theta_{m}$, and under the domain heterogeneity, the difference between the global and the clients becomes even more significant, which is quantified by:
\begin{equation}
    \mathrm{Var}_{m}(\theta_{m})=\frac{1}{M} \sum_{m=1}^{M} \bigg\lVert \theta_{g}-\theta_{m} \bigg\lVert.
\end{equation}
When aggregated at the server, the global model obtained from (\ref{FedAvg}) is far from any client's optimum, increasing the global empirical risk and leading to unstable convergence. In practice, the resulting model not only underperforms within each domain but also fails to achieve the objective of generalization across domains.
\vspace{-2mm}
\subsection{Global Representation Feature and Alignment Loss}
\begin{figure*}[t]
\centering
\includegraphics[width=0.88\textwidth]{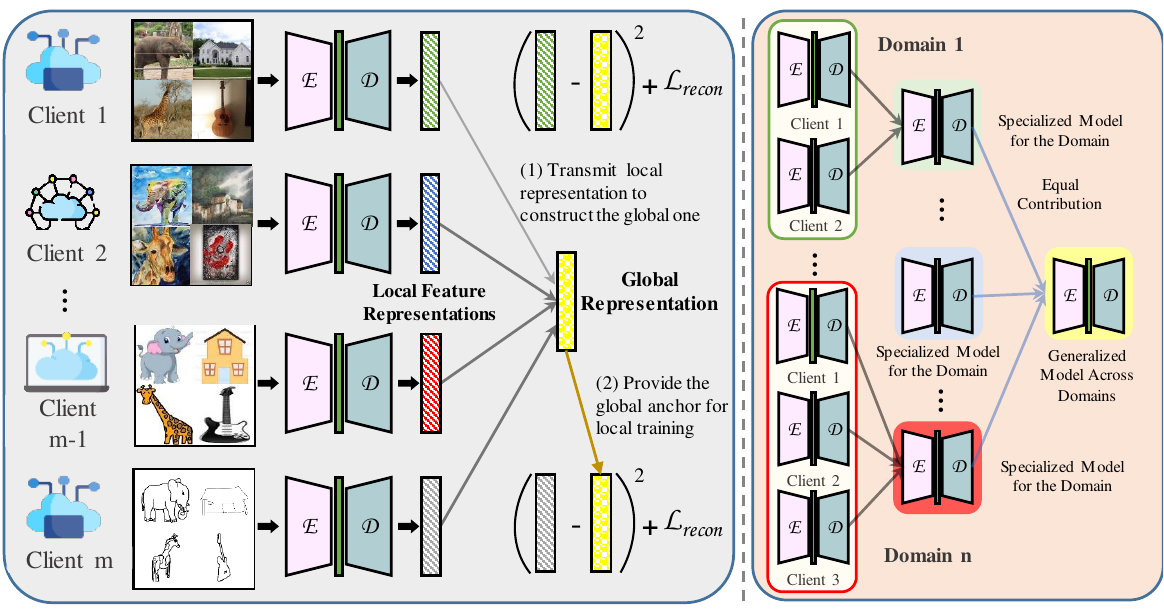}
\caption{(LHS) The local feature representation extractions and the construction of global representations; (RHS) The domain-aware aggregation approach to overcome the dominance of a subset of clients.}
\label{Proposal}
\end{figure*}

To overcome the discrepancy in the updated parameter directions across clients' learning models from different domains, we construct a global representation feature at the server by combining the local feature domains from clients, as shown in Fig.~\ref{Proposal}. Therefore, this representation contains knowledge about the clients' domain data and can be integrated into local training, thereby preventing the drift of local parameters. We demonstrate the construction process as below. Under the training process of each client, we extract the local feature representation of the client data as follows:
\begin{equation}
    F^{S}_{m}= \frac{1}{D_{m}}\sum_{i=1}^{D_{m}} f_{\theta} (I_{i})\in \mathbb{R}^{C},
\end{equation}
where $f_{\theta}(\cdot)$ is the semantic communication model of the client $m$, and $F^{S}$ is the summarization of the model output for all images before the reshaping process over the total number of client data. The summarization and averaging ensure the privacy of the training client, allowing the local feature representation to be transmitted safely to the server. In addition, summarizing all client data ensures the local feature captures a generalization of the domain data, thereby better illustrating the domain's characteristics. On the server side, we construct the global representation by averaging the local features from clients, which helps it capture the hidden latent structure of local data from different source domains. The mathematical equation for the global representation is provided as follows:
\begin{equation}
    G= \frac{1}{M}\sum_{m=1}^{M} F_{m}^{S}, G \in \mathbb{R}^{C}.
\end{equation}

After the construction of the global representation, we deliver it to the client side for the training. Specifically, we design a generalization loss $\mathcal{L}_{G}$ at the client side as follows:
\begin{equation}
    \mathcal{L}_{G}= \mathrm{MSE}(G,F_{m}^{S}).\label{globallosss}
\end{equation}
This generalization loss aligns local data features of a specific domain $S_{m}$ with the global representation across all domains. Compared to a scenario where the local model is trained with a single loss value, the generalization loss provides a soft constraint on parameter updates, inspired by device-level systems constraint in \cite{li2020federated}. To be specific, when the local model's parameters are trained toward the optimal value for its own domain data, it loses generalization capacity, the generalization loss becomes significantly large, and it will steer the learning process toward a more balanced position. Our loss is combined with the loss of the image reconstruction task to achieve high performance for the general data, regardless of the domain as follows:
\begin{equation}
    \mathcal{L}_{\mathrm{tot}} = \mathcal{L}_{\mathrm{recon}} + \lambda\cdot \mathcal{L}_{G},\label{finalloss}
\end{equation}
where $\lambda$ represents the contribution weight of the generalization loss into the client learning process. 
\subsection{Domain-aware Aggregation for Global Model}
Secondly, we identify the dominance problem in the global model aggregation process of the conventional FL framework, where a client with a large amount of data in a specific domain pushes the global model toward its own distribution \cite{mohri2019agnostic}. In FedAvg-like algorithms, the global parameters are updated as a weighted average of local models, where the weights correspond to the data sizes of each client. As a result, clients possessing more abundant data dominate the aggregation, overshadowing the contribution of others. It is worth noting that in real-world scenarios, data from different domains can vary greatly in both availability and complexity. For instance, an artist may require hours or even months to produce a single artwork, while sketch images can be generated within minutes and in large quantities. Consequently, the server may overfit to the easily generated but less semantically rich domain, compromising the generalization of the global model to more meaningful yet underrepresented domains. This imbalance in data samples across domains underscores the necessity of a domain-aware aggregation strategy that accounts semantic alignment and domain diversity rather than solely relying on data volume. To address this domain dominance, we propose a simple yet effective domain-aware mechanism to improve the model's generalization performance. Specifically, on the server, we first aggregate the models from clients in the same domain to build a specialized model for that domain, utilizing domain-specific information that is general and can be shared with the server. We have a total of four domains: photo, art painting, cartoon, and finally sketch, as shown in Fig.~\ref{Proposal}. The specialized model for the domain $u$ is presented as follows:
\vspace{-2mm}
\begin{equation}
    \theta^{r+1}_{u,g}=\sum_{m\in M, u} \frac{D_{m}}{D_{u}}\cdot\theta_{m}^{r+1},\label{domainspecializedmodels}
\end{equation}
where $\theta_{u,g}(\cdot)$ denotes the global model for the domain $u$, and $D_{u}$ is the total amount of data across clients from the corresponding domain $u$. Then, we aggregate the final models from the specialized models across all domains, giving each domain equal contributions, which ensures that each domain's data is treated fairly and not favored over any particular one.
\vspace{-2mm}
\begin{equation}
    \theta^{r+1}_{g}=\frac{1}{n}\sum_{u=1}^{n}\theta^{r+1}_{u,g}.
\end{equation}

With our proposed aggregation approach, clients with a large amount of data can contribute a greater portion of their trained weights to the specialized models for their corresponding domain, which is reasonable because they are trained on a large amount of data and thus have more advanced models. On the other hand, our approach puts constraints on the weighted contribution of clients from one domain toward a different domain, which prevents the dominance of a high-sample domain in the global aggregation process\cite{wang2024taming}. This ensures the global model is fair to all domains and provides the system's generalization performance, as indicated in~\ref{dominancedomaindiscussion}. The sequential detail of the Cross-domain Federated learning framework is given in Algorithm~\ref{alg:Alg1}. In the first round, we only transmit the initial global model. 
\vspace{-4.5mm}
\subsection{Convergence Analysis}
\subsubsection{Assumptions} We follow three standard assumptions \cite{li2020federated}: \textit{1) Each local objective function is $L_{1}$-Lipschitz smooth and continuous}, \textit{2) the stochastic gradient $g_{m,r}=\nabla \mathcal{L}(\theta_{m,r},\xi_{r})$ is unbiased and its variance bounded by $\sigma^{2}$}, \textit{3) the expected stochastic gradient is bounded by $V$}. On the other hand, we consider the fourth assumption for local representation extraction function, being $L_{2}$-Lipschitz continuous, which can be represented as the following equation\cite{tan2022fedproto}:
\vspace{-1.5mm}
\begin{equation}
    \lVert f_{\theta_{m,r_1}} (I) - f_{\theta_{m,r_2}} (I) \leq L_{2}  \lVert  \theta_{m,r_1}-\theta_{m,r_2}\lVert_{2}, \forall r_1, r_2, m\in M.  
\end{equation}
Given the assumptions above, we establish the theoretical convergence analysis, with Theorem 1 detailing the expected decrease in each round, where $e$ represents the local iteration and $R$ represents the global round.\\
\textbf{Theorem 1.} \textit{For an arbitrary client, after every communication round, we have,}
\vspace{-3mm}
\begin{align}
\mathbb{E}[\mathcal{L}_{(r+1)E + \frac{1}{2}}]
\leq\; &\mathcal{L}_{rE+\frac{1}{2}}
- \Big( \eta - \frac{\mathcal{L}_{1}\eta^{2}}{2} \Big)
    \sum_{e=\frac{1}{2}}^{E-1} \lVert \nabla\mathcal{L}_{rE+e} \rVert^{2}_{2} \nonumber\\
&\quad + \frac{L_{1}E\eta^{2}}{2}\sigma^{2}
+ \lambda L_{2}\eta EV,
\end{align}
\textit{where $rE$ is the timestep before the global feature representation. $rE+\frac{1}{2}$ is the time step right after the global representation and right before its first local step. The theorem indicates the deviation bound of the local objective function after each communication round, and convergence can be assured if a certain expected one-round decrease (with appropriate $\eta$ and $\lambda$). The sketch of proof is as follows:}\\

\textbf{Lemma 1.} \textit{With the assumptions 1 and 2 hold, the loss function of a client for communication round $r+1$ to the last local update step can be bounded as:}
\vspace{-0.5mm}
\begin{align}
\scalebox{0.93}{$
    \mathbb{E}[\mathcal{L}_{(r+1)E}] \leq \mathcal{L}_{rE+\frac{1}{2}}-(\eta-\frac{L_{1}\eta^{2}}{2})\sum_{e=\frac{1}{2}}^{E-1}\lVert \nabla\mathcal{L}_{rE+e}\lVert^{2}_{2} +\frac{L_{1}E\eta^{2}}{2}\sigma^2.$}
\end{align}

\textbf{Proof:} \textit{With the $\theta_{rE+1}= \theta_{rE+\frac{1}{2}}-\eta g_{rE +\frac{1}{2}}$, Assumption 1 can be written:} 
\begin{align}
\scalebox{0.94}{$
    \mathcal{L}_{rE+1} \leq \mathcal{L}_{rE+\frac{1}{2}} -\eta \langle\nabla \mathcal{L}_{rE+\frac{1}{2}}, g_{rE+\frac{1}{2}}\rangle + \frac{L_{1}}{2}\lVert \eta g_{rE+\frac{1}{2}} \lVert^{2}_{2}.$}
\end{align}
\textit{We take the expectation of both sides on the random variable $\xi_{rE+\frac{1}{2}}$, which we obtain:}
\vspace{-1mm}
\begin{align}
    \mathbb{E}[\mathcal{L}_{rE+1}] \leq \mathcal{L}_{rE+\frac{1}{2}} - \eta \mathbb{E}[\langle\nabla \mathcal{L}_{rE+\frac{1}{2}}, g_{rE+\frac{1}{2}}\rangle] &
    \nonumber &\\+ \frac{L_{1}\eta^2}{2} \mathbb{E}[\lVert  g_{rE+\frac{1}{2}} \lVert^{2}_{2}],
\end{align}
\textit{which we then consider the unbiased stochastic gradient from Assumption 2 }($\mathbb{E}_{\xi_{m}\sim D_{m}}= \nabla \mathcal{L}(\theta_{m,r}) = \nabla\mathcal{L}_{r}$), $Var(x)= \mathbb{E}[x^2]-(\mathbb{E}[x])^2$, \textit{and finally the variance bounded of Assumption 2} ($\mathbb{E}[\lVert g_{m,r} -\nabla \mathcal{L}(\theta_{m,r})\lVert^{2}_{2}] \leq \sigma^{2}$) \textit{, which result in:}
\begin{align}
\scalebox{0.96}{$
    \mathbb{E}[\mathcal{L}_{rE+1}] \leq  \mathcal{L}_{rE+\frac{1}{2}} - (\eta - \frac{L_{1}\eta^{2}}{2}) \lVert\nabla \mathcal{L}_{rE+\frac{1}{2}}\lVert_{2}^2 + \frac{L_{1}\eta^{2}}{2}\sigma^2. $}
\end{align}
\textit{Finally, we telescope of E iterations to derive Lemma 1.}

\textbf{Lemma 2.} \textit{Consider the assumptions 3 and 4 hold, after the global representation aggregation at the server, the loss function for a random client can be bounded as:}
\begin{equation}
    \mathbb{E}[\mathcal{L}_{(r+1)E+\frac{1}{2}}] \leq \mathcal{L}_{(r+1)E} + \lambda L_{2}\eta E V.
\end{equation}
\textbf{Proof:} \textit{Starting with our proposed local loss:}
\begin{align}
    \mathcal{L}_{(r+1)E+\frac{1}{2}} & = \mathcal{L}_{(r+1)E} + \lambda \lVert f({\theta}_{(r+1)E}) - G_{r+2} \lVert_{2} \nonumber \\&  - \lambda \lVert f(\theta_{(r+1)E})-G_{r+1}\lVert_{2} \label{lem2eq2},
\end{align}
\textit{we adopt the result of the following math} $\lVert a-b\lVert_{2}$-$\lVert a-c\lVert_{2} \leq $ $\lVert b-c\lVert_{2}$ \textit{to obtain} $\mathcal{L}_{(r+1)E+\frac{1}{2}} \leq \mathcal{L}_{(r+1)E} + \lambda \lVert G_{r+2}-G_{r+1}\lVert_{2}$. \textit{After which, we replace $G$ with the definition of global representation, $F$ local representation into the equation:}
\begin{align}
    \mathcal{L}_{(r+1)E+\frac{1}{2}} & \leq \mathcal{L}_{(r+1)E} + \lambda\lVert \sum_{m=1}^{M} \frac{1}{M D_{m}}\sum_{i=1}^{D_{m}}(f_{m}(\theta_{m,(r+1)E},I_{i})\nonumber  \\ &-f_{m}(\theta_{m,rE},I_{i}) \lVert_{2}\label{lem2eq6},
\end{align}
\textit{then we apply the math} $\lVert\sum a_{i}\lVert_{2} \leq $ $\sum \lVert a_{i}\lVert_{2}$ \textit{to obtain the} $\lVert (f_{m}(\theta_{m,(r+1)E},I_{i}) -f_{m}(\theta_{m,rE},I_{i}) \lVert_{2}$\textit{, whose values $\leq L_{2} \lVert\theta_{m,(r+1)E}-\theta_{m,rE}\lVert_{2}$ (Assumption 4), thus we obtain:}
\begin{align}
\scalebox{1}{$
    \mathcal{L}_{(r+1)E+\frac{1}{2}}  \leq \mathcal{L}_{(r+1)E} + \lambda L_{2} \eta\sum_{m=1}^{M}\frac{1}{M}\sum_{e=\frac{1}{2}}^{E-1}\lVert g_{m,rE+e}\lVert_{2}, $} \label{lem2eq10}
\end{align}
\textit{Then we take expectation of the random variable $\xi$:}
\begin{align}
    \mathbb{E}[\mathcal{L}_{(r+1)E+\frac{1}{2}}] & \leq  \mathcal{L}_{(r+1)E} + \lambda L_{2} \eta \sum_{m}^{M}\frac{1}{M}\sum_{e=\frac{1}{2}}^{E-1}\mathbb{E}[\lVert g_{m,rE+e}\lVert_{2}] \label{lem2eq11}\\& 
    \leq \mathcal{L}_{(r+1)E} +\lambda L_{2} \eta E V,\label{lem2eq12}
\end{align}
\textit{The transition from (\ref{lem2eq11}) to (\ref{lem2eq12}) follows directly from Assumption 3 and (\ref{lem2eq12}) establishes the result stated in Lemma 2. Finally, we take the expectation of $\theta$ on both sides in Lemmas 1 and 2, and then sum them up to derive Theorem 1.}

\textbf{Corollary 1.} \textit{The loss function $\mathcal{L}$ of an arbitrary client monotonously decreases in every communication round when $-(\eta-\frac{L_{1}\eta^2}{2})\sum_{e=1/2}^{E-1}\lVert\nabla\mathcal{L}_{rE+e}\lVert_{2}^{2}+\frac{L_{1}E\eta^2}{2}\sigma^2+\lambda L_{2} E V < 0$. Therefore, the value of learning rate $\eta$ is bounded}:
\begin{equation}
    \eta_{e'} < \frac{2\big(\sum_{e=\frac{1}{2}}^{e'}\lVert\mathcal{L}_{rE+e}\lVert^{2}_{2}-\lambda L_{2}E V\big)}{L_{1}\big(\sum_{e=\frac{1}{2}}^{e'}\lVert \nabla \mathcal{L}_{rE+e}\lVert^{2}_{2}+E\sigma^2\big)}, 
\end{equation}
\textit{where $e'$ runs from the first iteration step till the last batch of data, and}
\begin{equation}
    \lambda_{e}=\frac{\lVert \nabla\mathcal{L}_{rE+\frac{1}{2}}\lVert^{2}_{2}}{L_{2}EV}.
\end{equation}
\textit{Thus, the loss function converges.} 

\begin{algorithm}[t]
   \caption{\strut Cross-domain Federated Learning Framework for Semantic Communication} 
   \label{alg:Alg1}
   \begin{algorithmic}[1]
       \State{\textbf{Initialize:} Global model $\boldsymbol{\theta}$, Global Training Round $R$, number of gateways $G$.}
        \For{one global round r$=1,2,\ldots,R$}
        \State Broadcast the global model \& global feature to clients.
          \For{each client device $m\in M$ $\textbf{in parallel}$}
              \For{each step $e$ $\in$ $E$}
              \State{Feed forward the data batch into the model to}
              \NoNumber{obtain the reconstruction images and average}
              \NoNumber{feature representation.}
              \State{$F^{s}_{m}$=$F^{s}_{m}$+\big($f_{\theta^{r,e}_{m}}(I_{B})$\big)$/$$D_{m}$}
              \State{$\boldsymbol{\theta}^{r,e+1}_{m} \leftarrow \boldsymbol{\theta}^{r,e}_{m}- \eta \nabla \mathcal{L}^{r,e}_{tot}$.}
              \EndFor
        \State{Achieve the final model $\boldsymbol{\theta}_{m}^{r+1}$ \& the local feature}
        \NoNumber{representation $F_{m}^{r+1}$ of clients.}
            \EndFor
        \State{Sequentially aggregate the domain specialized models}
        \NoNumber{$\boldsymbol{\theta}^{r+1}_{u,g}$ with  (\ref{domainspecializedmodels}) and the new global model at the server,} 
        \NoNumber{along with the global representation.}
        \EndFor
    \State{\textbf{Output:} Global Model $\boldsymbol{\theta}$.}
   \end{algorithmic}
\end{algorithm}
\subsection{Swin Transformer-based DeepJSCC}
In this section, we provide the details of the DL architecture of the semantic communication model. Specifically, we implement the Swin Transformer model\cite{liu2021swin}, which has high capability in semantic extraction and also exhibits high computing efficiency, for the source encoder/decoder. Instead of executing a global attention mechanism over all the patches within the image, like the conventional Vision Transformer, the Swin model partitions the image into non-overlapping local windows and applies the attention mechanism to each window. Benefiting from this local self-attention, we can significantly reduce the computing operation by processing on a smaller region. To capture the long-range dependency correlation across windows, Swin Transformer proposes a shifted window operation to provide connections among windows. With an image having $h\times w $ patches, and each window contains $M \times M$ patches, the comparison of computation complexity between the global multi-head self-attention (MSA) and the window-based is presented according to \cite{liu2021swin}
    \begin{equation}
        \Omega(\textrm{MSA})= 4hwC^{2} + 2(hw)^2C,
    \end{equation}
    \begin{equation}
        \Omega(\textrm{W-MSA})= 4hwC^{2} + 2M^2hwC.
    \end{equation}

The \textrm{MSA} scales quadratically with the number of patches $hw$, and \textrm{W-MSA} only scales linearly with the image size, while we can set the size of $M$. Specifically, our encoder model consists of two Swin Transformer stages, where the first stage comprises two Transformer blocks, and the subsequent stage features four blocks. On the other hand, the source decoder has a symmetric architecture toward the source encoder, where the first stage has four Transformer blocks and the second has two. The symmetric architecture between the source encoder and decoder enhances the capability for signal decoding, thereby improving end-to-end performance. On the other hand, we adopt a simple yet effective DL architecture \cite{DJSCCSwin} for the channel coders to compress and decompress the semantic features. Specifically, it consists of a set of fully connected layers combined with an SNR fusion mechanism, which enables the model to adapt its encoding and decoding process according to the current channel condition, thereby improving robustness and reconstruction fidelity under varying noise levels.
\vspace{-3mm}
\section{Simulation Results and Analysis}\label{Performance}

Within this section, we conduct a comprehensive series of simulations to effectively demonstrate the outstanding performance of our frameworks when the client's data originates from various data sources. First, we provide the simulation setting for our training framework and also the communication environment for the semantic communication system. Secondly, we present benchmark schemes for comparison analysis, and finally, we evaluate system performance under our proposed approaches in comparison to others. We refer to our framework as FedDoM.

\subsection{Simulation Configuration for the FL \& the Semantic Communication Model}
\subsubsection{Datasets, Distributions, and DL-based architecture} We adopt the PACS dataset for our training and evaluation process due to its diverse image domains, including photo, sketch, art painting, and cartoon\cite{li2017deeper}. Similar to other studies, we sample $30\%$ of the full-size datasets for each domain, and we divide them into $80\%$ for training process and $20\%$ for testing. Then, we partition these data samples into a total of ten clients using the Dirichlet distribution with $\alpha=1$. Specifically, for each domain, the data is divided into two or three clients, and each client is limited to one type of domain data, which exacerbates the heterogeneity in domains. We provide the number of data samples of clients and their domains as in Table~\ref{trainingdata}.

Unlike other conventional FL approaches that focus on classification tasks, the objective of our semantic communication model is image reconstruction, which requires a more complex DL architecture. Therefore, we adopt the Swin-Transformer model, which has high capability in semantic extraction and low computational resource requirements, for the implementation of DL-based semantic communication.

We apply the power normalization operation to the output of the DL-based JSCC encoder before transmitting it, as described in \cite{DJSCCnofeedback}. To simulate the dynamic of channel noise, we randomly sample a value from the set of $\mathcal{S}$ $\in \{1, 3, 5, 7, 9\}$, while the transmit power is fixed at $1$ for all the experiments. The learning rate $\eta$ of DL models is $1e^{-3}$, the training batch size is $16$, and the total global training rounds are $500$.
\subsubsection{Benchmarks}
To demonstrate the performance gain of our proposal framework, we conduct the simulation experiments against several classical benchmarks as follows:
\begin{itemize}
    \item FedAvg: It is a traditional federated learning framework, where clients are trained independently of each other, and the BS aggregates the global model using the proportion of the number of data samples from each client\cite{mcmahan2017communication}.
    \item FedProx: It introduces a regularization term to limit the deviation of each local model from the global model in the local training process, thereby helping to mitigate data heterogeneity among clients in the network \cite{li2020federated}.
    \item MOON: It proposes a contrastive loss that encourages each client’s model to stay closer to the global model and farther from outdated local models, helping align representations across clients\cite{li2021model}.
\end{itemize}

\subsection{Domain Shift in Semantic Communication Model}
\begin{figure*}[t]
\centering
\includegraphics[width=0.90\textwidth]{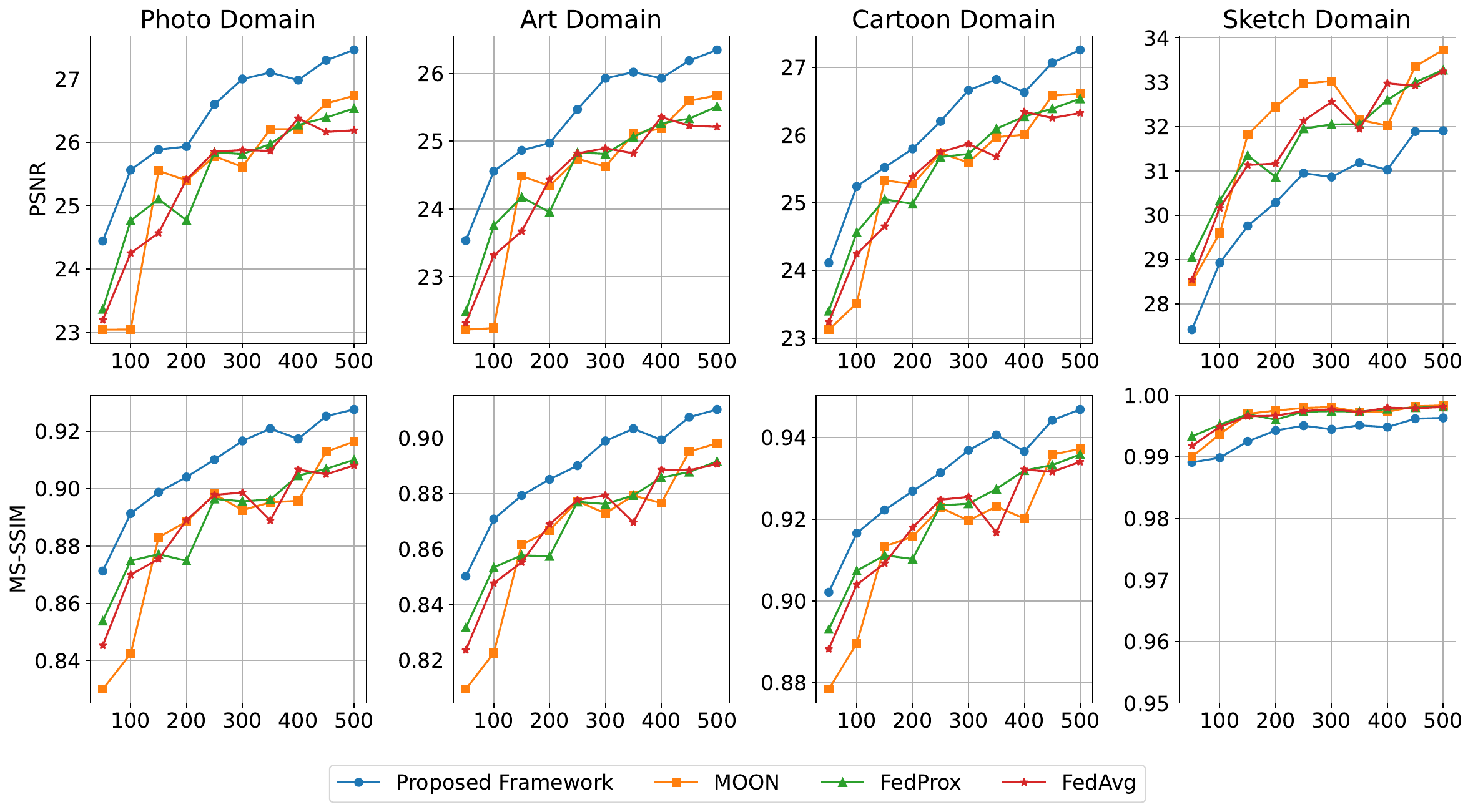}
\caption{The comprehensive performance of the proposed frameworks against other FL approaches for four kinds of domain data. The comparison performance is measured in terms of PSNR and MS-SSIM metrics.}
\label{Convergence}
\end{figure*}

\begin{table}[t]
\centering
\caption{Number of training sample for clients and its corresponding domains}
\renewcommand{\arraystretch}{1.0} 
\scalebox{1.05}{
\begin{tabular}{|c|c|c|c|c|c|}
\hline
Domain                   & Client & Samples & Domain                  & Client & Samples \\ \hline
\multirow{2}{*}{Photo}   & \#1    & 43           & \multirow{3}{*}{Art}    & \#6    & 70           \\ \cline{2-3} \cline{5-6} 
                         & \#2    & 358          &                         & \#7    & 303          \\ \cline{1-3} \cline{5-6} 
\multirow{3}{*}{Cartoon} & \#3    & 87           &                         & \#8    & 119          \\ \cline{2-6} 
                         & \#4    & 93           & \multirow{2}{*}{Sketch} & \#9    & 827          \\ \cline{2-3} \cline{5-6} 
                         & \#5    & 383          &                         & \#10   & 116          \\ \hline
\end{tabular}}
\label{trainingdata}
\end{table}

\begin{table}[t]
\centering
\caption{average performance for all the domains}
\renewcommand{\arraystretch}{1.} 
\scalebox{1.05}{
\begin{tabular}{|c|c|c|c|c|}
\hline
Frameworks & Proposal & MOON    & FedProx & FedAvg  \\ \hline
PSNR    & \textbf{28.2413}  & 28.1862 & 27.9640 & 27.7421 \\ \hline
MS-SSIM & \textbf{0.9452}   & 0.9375  & 0.9339  & 0.9327  \\ \hline
\end{tabular}}
\label{AveragePerformance}
\end{table}
\subsubsection{Performance Convergence}
\begin{table*}[t]
\centering
\caption{Performance of the Semantic Communication System being trained under different FL frameworks}
\renewcommand{\arraystretch}{1.05} 
\scalebox{1.05}{
\begin{tabular}{|c|c|cc|cc|cc|cc|}
\hline
\multirow{2}{*}{}         AWGN      & \multirow{2}{*}{\begin{tabular}[c]{@{}c@{}}Channel\\ Condition SNR\end{tabular}} & \multicolumn{2}{c|}{FedAvg}                    & \multicolumn{2}{c|}{FedProx}           & \multicolumn{2}{c|}{MOON}                               & \multicolumn{2}{c|}{FedDoM (Ours)}                             \\ \cline{3-10} 
                                &                                                                        & \multicolumn{1}{c|}{PSNR}    & MS-SSIM         & \multicolumn{1}{c|}{PSNR}    & MS-SSIM & \multicolumn{1}{c|}{PSNR}             & MS-SSIM         & \multicolumn{1}{c|}{PSNR}             & MS-SSIM         \\ \hline
\multirow{4}{*}{Photo Domain}   & 1 dB                                                                   & \multicolumn{1}{c|}{23.9898} & 0.8395          & \multicolumn{1}{c|}{24.3577} & 0.8430  & \multicolumn{1}{c|}{24.5203}          & 0.8530          & \multicolumn{1}{c|}{\textbf{25.2305}} & \textbf{0.8703} \\ \cline{2-10} 
                                & 4 dB                                                                   & \multicolumn{1}{c|}{25.7241} & 0.8949          & \multicolumn{1}{c|}{26.0667} & 0.8971  & \multicolumn{1}{c|}{26.2694}          & 0.9045          & \multicolumn{1}{c|}{\textbf{26.9768}} & \textbf{0.9170} \\ \cline{2-10} 
                                & 7 dB                                                                   & \multicolumn{1}{c|}{26.9677} & 0.9280          & \multicolumn{1}{c|}{27.3434} & 0.9301  & \multicolumn{1}{c|}{27.5417}          & 0.9346          & \multicolumn{1}{c|}{\textbf{28.2844}} & \textbf{0.9437} \\ \cline{2-10} 
                                & 10 dB                                                                  & \multicolumn{1}{c|}{27.7098} & 0.9446          & \multicolumn{1}{c|}{28.1969} & 0.9480  & \multicolumn{1}{c|}{28.3099}          & 0.9510          & \multicolumn{1}{c|}{\textbf{29.1573}} & \textbf{0.9580} \\ \hline
\multirow{4}{*}{Art Domain}     & 1 dB                                                                   & \multicolumn{1}{c|}{22.9119} & 0.8119          & \multicolumn{1}{c|}{23.2517} & 0.8149  & \multicolumn{1}{c|}{23.3539}          & 0.8251          & \multicolumn{1}{c|}{\textbf{24.0118}} & \textbf{0.8421} \\ \cline{2-10} 
                                & 4 dB                                                                   & \multicolumn{1}{c|}{24.7294} & 0.8759          & \multicolumn{1}{c|}{26.3322} & 0.8770  & \multicolumn{1}{c|}{25.1907}          & 0.8847          & \multicolumn{1}{c|}{\textbf{25.8448}} & \textbf{0.8978} \\ \cline{2-10} 
                                & 7 dB                                                                   & \multicolumn{1}{c|}{26.0138} & 0.9126          & \multicolumn{1}{c|}{26.3321} & 0.9137  & \multicolumn{1}{c|}{26.4754}          & 0.9183          & \multicolumn{1}{c|}{\textbf{27.1921}} & \textbf{0.9289} \\ \cline{2-10} 
                                & 10 dB                                                                  & \multicolumn{1}{c|}{26.8009} & 0.9311          & \multicolumn{1}{c|}{27.2026} & 0.9336  & \multicolumn{1}{c|}{27.3012}          & 0.9366          & \multicolumn{1}{c|}{\textbf{28.0983}} & \textbf{0.9455} \\ \hline
\multirow{4}{*}{Cartoon Domain} & 1 dB                                                                   & \multicolumn{1}{c|}{24.0187} & 0.8818          & \multicolumn{1}{c|}{24.1687} & 0.8833  & \multicolumn{1}{c|}{24.2266}          & 0.8854          & \multicolumn{1}{c|}{\textbf{24.7267}} & \textbf{0.8978} \\ \cline{2-10} 
                                & 4 dB                                                                   & \multicolumn{1}{c|}{25.8879} & 0.9249          & \multicolumn{1}{c|}{26.0961} & 0.9270  & \multicolumn{1}{c|}{26.1705}          & 0.9286          & \multicolumn{1}{c|}{\textbf{26.7655}} & \textbf{0.9386} \\ \cline{2-10} 
                                & 7 dB                                                                   & \multicolumn{1}{c|}{27.0243} & 0.9481          & \multicolumn{1}{c|}{27.2331} & 0.9494  & \multicolumn{1}{c|}{27.2992}          & 0.9500          & \multicolumn{1}{c|}{\textbf{28.0201}} & \textbf{0.9588} \\ \cline{2-10} 
                                & 10 dB                                                                  & \multicolumn{1}{c|}{27.7154} & 0.9607          & \multicolumn{1}{c|}{27.9330} & 0.9621  & \multicolumn{1}{c|}{27.9693}          & 0.9616          & \multicolumn{1}{c|}{\textbf{28.7795}} & \textbf{0.9695} \\ \hline
\multirow{4}{*}{Sketch Domain}  & 1 dB                                                                   & \multicolumn{1}{c|}{30.9371} & \textbf{0.9949} & \multicolumn{1}{c|}{30.9542} & 0.9945  & \multicolumn{1}{c|}{\textbf{31.3318}} & 0.9947          & \multicolumn{1}{c|}{29.8258}          & 0.9912          \\ \cline{2-10} 
                                & 4 dB                                                                   & \multicolumn{1}{c|}{32.7815} & 0.9978          & \multicolumn{1}{c|}{32.8184} & 0.9978  & \multicolumn{1}{c|}{\textbf{33.2617}} & \textbf{0.9981} & \multicolumn{1}{c|}{31.5196}          & 0.9958          \\ \cline{2-10} 
                                & 7 dB                                                                   & \multicolumn{1}{c|}{33.9914} & 0.9986          & \multicolumn{1}{c|}{34.0131} & 0.9986  & \multicolumn{1}{c|}{\textbf{34.4753}} & \textbf{0.9988} & \multicolumn{1}{c|}{32.5133}          & 0.9971          \\ \cline{2-10} 
                                & 10 dB                                                                  & \multicolumn{1}{c|}{34.7360} & 0.9990          & \multicolumn{1}{c|}{34.7430} & 0.9990  & \multicolumn{1}{c|}{\textbf{35.2475}} & \textbf{0.9991} & \multicolumn{1}{c|}{33.1311}          & 0.9978          \\ \hline
\end{tabular}}
\label{PerformanceAWGN}
\end{table*}
\begin{figure*}[t]
\centering
\includegraphics[width=0.9\textwidth]{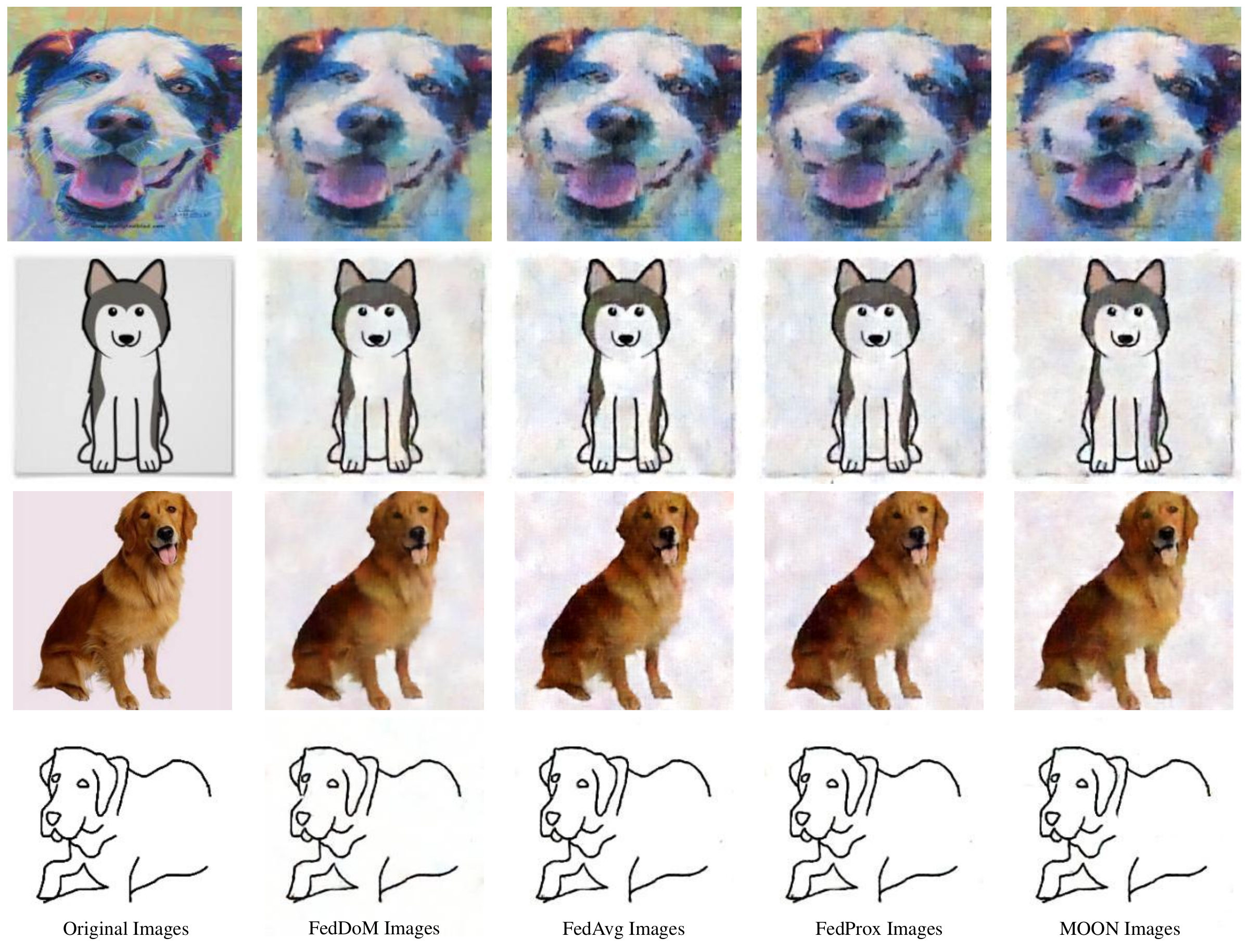}
\caption{Visualization of the original images and the reconstructed outputs produced by FedDoM, FedAvg, FedProx, and MOON. Notable differences appear particularly in the background regions of the Cartoon and Photo samples, where FedDoM yields smoother textures and reduced noise compared with the other frameworks.}
\label{Visual}
\end{figure*}

As shown in Fig.~\ref{Convergence}, we present the convergence performance of our training framework compared to the benchmarks, with an SNR of $5$ dB. In general, the proposed framework reaches high PSNR and MS-SSIM values in fewer rounds than MOON, FedProx, and FedAvg, showing both rapid error reduction and early capture of salient semantics. Specifically, we achieve higher results for the Photo, Art, and Cartoon domains, whose data structures are more complex than those of the Sketch domain, which is relatively simple. Across different domains, our curves have fewer sudden jumps and shorter transients than the baselines, showing more consistent updates each round. Within this section, we focus on the convergence curves of the frameworks, and later, we investigate the reason for the less impressive performance for sketch data and explain why such behavior is desirable. Furthermore, we provide the average performance across domains in Table~\ref{AveragePerformance}, where the proposed system achieves the highest performance for both metrics. This outstanding performance effectively demonstrates the generality of the semantic communication system for handling data from diverse sources.
\begin{table*}[t]
\centering
\caption{sensitivity analysis of the scaling coefficient $\lambda$ on PSNR and MS-SSIM under Different Domains and Channel Conditions}
\renewcommand{\arraystretch}{1.05}
\scalebox{1.05}{
\begin{tabular}{|c|c|cccc|cccc|}
\hline
\multirow{2}{*}{$\lambda$} & \multirow{2}{*}{Domain} & \multicolumn{4}{c|}{PSNR (dB)} & \multicolumn{4}{c|}{MS-SSIM} \\ \cline{3-10}
 &  & 1 dB & 4 dB & 7 dB & 10 dB & 1 dB & 4 dB & 7 dB & 10 dB \\ \hline
\multirow{4}{*}{1.0} 
 & Art     & 24.0118 & 25.8448 & 27.1921 & 28.0983 & 0.8421 & 0.8978 & 0.9289 & 0.9455 \\
 & Cartoon & 24.7267 & 26.7655 & 28.0201 & 28.7795 & 0.8978 & 0.9386 & 0.9588 & 0.9695 \\
 & Photo   & 25.2305 & 26.9768 & 28.2844 & 29.1573 & 0.8703 & 0.9170 & 0.9437 & 0.9580 \\
 & Sketch  & 29.8258 & 31.5196 & 32.5133 & 33.1311 & 0.9912 & 0.9958 & 0.9971 & 0.9978 \\ \hline
\multirow{4}{*}{1.5} 
 & Art     & 24.0083 & 25.8606 & 27.2360 & 28.1643 & 0.8415 & 0.8979 & 0.9295 & 0.9466 \\
 & Cartoon & 24.7173 & 26.7943 & 28.0393 & 28.8091 & 0.8977 & 0.9384 & 0.9584 & 0.9695 \\
 & Photo   & 25.1902 & 26.9673 & 28.3059 & 29.2009 & 0.8687 & 0.9165 & 0.9439 & 0.9588 \\
 & Sketch  & 29.9034 & 31.5663 & 32.5568 & 33.1539 & 0.9919 & 0.9961 & 0.9973 & 0.9979 \\ \hline
\multirow{4}{*}{2.0} 
 & Art     & 23.8546 & 25.6333 & 26.9086 & 27.7467 & 0.8395 & 0.8950 & 0.9262 & 0.9426 \\
 & Cartoon & 24.6644 & 26.6521 & 27.8366 & 28.5531 & 0.8961 & 0.9378 & 0.9576 & 0.9681 \\
 & Photo   & 25.0641 & 26.7486 & 27.9537 & 28.7487 & 0.8682 & 0.9150 & 0.9417 & 0.9559 \\
 & Sketch  & 30.0704 & 31.6931 & 32.6668 & 33.2482 & 0.9921 & 0.9963 & 0.9973 & 0.9978 \\ \hline
\end{tabular}}
\label{tab:lambda_sensitivity}
\end{table*}
\subsubsection{Performance under a wide range of Channel Conditions} In Table~\ref{PerformanceAWGN}, we present the PSNR and MS-SSIM metrics in all training frameworks for a comprehensive comparison. Overall, the FedDoM secures the DL-based semantic communication model with the most robust capability, followed by MOON, FedProx, and finally FedAvg. For the photo domain, the performance gap between the proposed FedDoM and the second-best approach, MOON, is $0.701$ and $0.017$ when the channel noise is at $1$ dB. In the meantime, our proposal outperforms the FedAvg and FedProx up $1.2407$ and $0.8728$ for the PSNR metric, $0.034$ and $0.027$ in terms of MS-SSIM. These performance gaps have an increasing trend when the quality of the wireless channel improves. 

We also encounter a similar pattern for the performance results for the Art and Cartoon domains. Specifically, the performance gap is around $0.6579$ at the $1$ dB and increases to $0.7971$ at $10$ dB, while there is a significant difference in MS-SSIM value between training approaches at $1$ dB,  which is around $0.017$; however, due to the maximum value of MS-SSIM is $1$, then the metric performance gap encounters the saturation point at $10$ dB with the value $0.0089$.

The Sketch domain exhibits a different pattern. Here, MOON attains the highest PSNR and MS-SSIM across SNRs, while FedDoM is outperformed. We attribute this to the relative simplicity and low semantic variability of sketch images: sketches contain sparse, high-contrast structure and fewer texture cues, so methods that strongly preserve local features (e.g., contrast- and edge-preserving objectives used in some domain-adaptation training) can produce superior numerical fidelity. In contrast, FedDoM’s domain-aware clustering and global-alignment mechanisms prioritize generalization across diverse, high-variation domains; this design yields superior performance on Photo, Art, and Cartoon images but can slightly reduce performance on low-variance domain (Sketch), where specialized, domain-focused optimization is most beneficial.

Taken together, these results confirm two key points. First, domain-aware aggregation combined with global representation alignment substantially improves cross-domain generalization for complex-image domains and benefits more from better channel conditions. Second, there is an inherent trade-off between cross-domain generality and per-domain specialization, demonstrating the need for mechanisms that adapt aggregation strength based on domain complexity or allow hybrid strategies (e.g., per-domain fine-tuning) when domain homogeneity is high.
\begin{figure*}[t]
\centering
\includegraphics[width=0.9\textwidth]{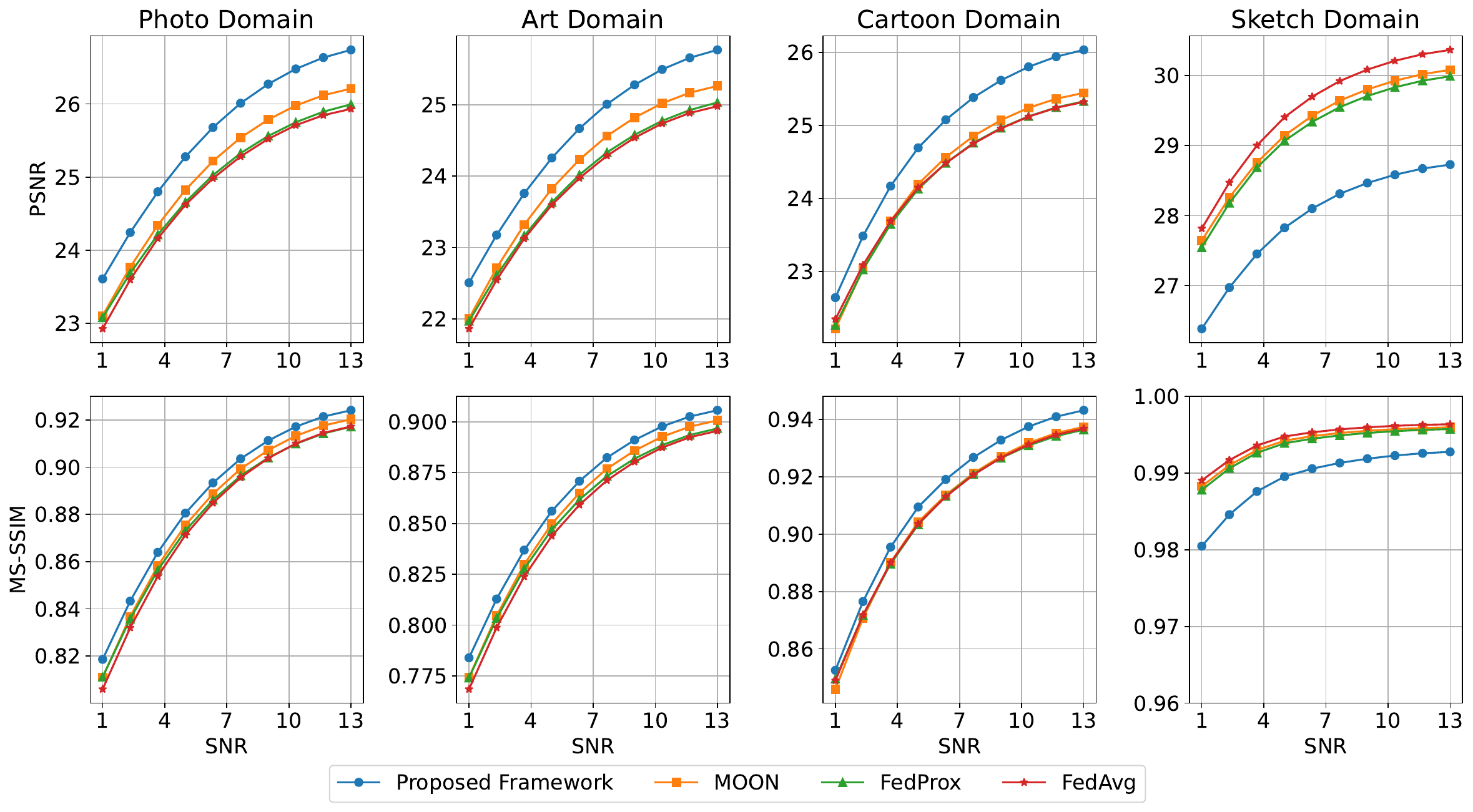}
\caption{The performance comparison of the proposed frameworks against other benchmarks under four domains: Photo, Art Painting, Cartoon, and finally Sketch. These results are obtained under a Rayleigh fading channel with an SNR range of 1 to 13dB.}
\label{rayleigh}
\end{figure*}
\subsubsection{Visual Performance}
In Fig.~\ref{Visual}, we qualitatively compare the reconstructed images obtained by different FL training frameworks under the same channel condition. It is evident that the proposed FedDoM framework produces reconstructions that are closest to the original images in terms of both structural integrity and perceptual detail. In particular, FedDoM effectively preserves fine-grained textures and semantic attributes-such as object boundaries, color consistency, and spatial coherence-while the baseline approaches (FedAvg, FedProx, and MOON) often exhibit blurring, color distortion, or loss of semantic content. These degradations are especially noticeable in the background of the Photo and cartoon domains, where our model achieves smoother results compared to the remaining benchmarks and also closely resembles the original image. In addition, FedDoM maintains consistent reconstruction quality across domains, demonstrating the benefit of the proposed global representation alignment and domain-aware aggregation mechanisms. Overall, this qualitative analysis aligns well with the quantitative results in Table~\ref{PerformanceAWGN}, confirming that FedDoM achieves superior cross-domain generalization and perceptual fidelity under various channel conditions.

\subsubsection{Sensitive analysis for the coefficient value of the global representation}

So far, we have conducted the experiment with the value of $\lambda$ equal to one, and in Table~\ref{tab:lambda_sensitivity}, we present the sensitivity analysis of the scaling coefficient $\lambda$ for the generalization alignment loss, evaluated across four domains under various SNR levels. Overall, the results indicate that the value of $\lambda$ has a notable influence on the reconstruction quality of the semantic communication system. When $\lambda = 1$, the model achieves high reconstruction performance, particularly for the Photo and Cartoon domains, suggesting that moderate weighting allows effective alignment between the global and local representations. Increasing $\lambda$ to $1.5$ yields slightly improved performance across most SNR levels, showing that a stronger emphasis on alignment encourages more stable feature learning and better generalization across domains. However, further increasing $\lambda$ to $2$ leads to a mild degradation in both PSNR and MS-SSIM, especially in the Art and Photo domains. This suggests that excessive alignment strength constrains domain-specific features and limits the model’s flexibility to adapt to distinct data characteristics.

\begin{table}[t]
\centering
\caption{data sample under high level of heterogeneity of domain data}
\renewcommand{\arraystretch}{1} 
\scalebox{1.05}{
\begin{tabular}{|c|c|c|c|c|c|}
\hline
Domain                   & Client & Samples & Domain                  & Client & Samples \\ \hline
\multirow{2}{*}{Photo}   & \#1    & 210           & \multirow{3}{*}{Art}    & \#6    & 223           \\ \cline{2-3} \cline{5-6} 
                         & \#2    & 58          &                         & \#7    & 85          \\ \cline{1-3} \cline{5-6} 
\multirow{3}{*}{Cartoon} & \#3    & 82           &                         & \#8    & 20          \\ \cline{2-6} 
                         & \#4    & 61           & \multirow{2}{*}{Sketch} & \#9    & 666          \\ \cline{2-3} \cline{5-6} 
                         & \#5    & 232          &                         & \#10   & 906          \\ \hline
\end{tabular}}
\label{unbalanced}
\end{table}

Across all settings, the Sketch domain exhibits consistently high MS-SSIM values with minor variation, indicating that its simpler structural patterns make it less sensitive to the scaling coefficient. In contrast, the Art and Photo domains, which contain more complex and diverse textures, are more affected by the choice of $\lambda$. These results confirm that setting $\lambda = 1.5$ offers the most balanced trade-off between cross-domain generalization and domain-specific fidelity, and thus it is adopted as the default configuration for subsequent experiments.

\begin{table*}[t]
\centering
\scriptsize
\caption{Comparison of PSNR and MS-SSIM Across Domains and FL Frameworks Under Extreme Data-Domain Heterogeneity.}
\renewcommand{\arraystretch}{1.05}
\scalebox{1.1}{
\begin{tabular}{|c|c|c|c|c|c|c|}
\hline
\textbf{SNR (dB)} & \textbf{Framework} 
& \textbf{Photo} & \textbf{Art Painting} & \textbf{Cartoon} & \textbf{Sketch} & \textbf{Avg of All} \\ \hline

\multirow{4}{*}{1} 
& FedAvg
& 20.0568 / 0.6851 & 20.2223 / 0.6766 & 21.8236 / 0.7973 & 31.5472 / 0.9952 
& 23.4125 / 0.7885 \\ \cline{2-7}

& FedProx
& 21.6571 / 0.7615 & 21.8228 / 0.7543 & 22.9906 / 0.8416 & 32.1310 / 0.9959
& 24.6504 / 0.8383 \\ \cline{2-7}

& MOON
& 22.1661 / 0.7857 & 22.2466 / 0.7753 & 23.3503 / 0.8548 & \textbf{32.6507 / 0.9969}
& 25.1034 / 0.8532 \\ \cline{2-7}

& FedDoM
& \textbf{23.9585 / 0.8419} & \textbf{24.0686 / 0.8300} & \textbf{24.7176 / 0.8900} & 30.1060 / 0.9914
& \textbf{25.7127 / 0.8883} \\ \hline

\multirow{4}{*}{4} 
& FedAvg
& 22.1221 / 0.7872 & 22.2034 / 0.7758 & 23.5246 / 0.8584 & 33.4756 / 0.9978
& 25.3314 / 0.8548 \\ \cline{2-7}

& FedProx
& 23.6223 / 0.8435 & 23.7363 / 0.8354 & 24.7695 / 0.8950 & 34.1270 / 0.9983
& 26.5638 / 0.8930 \\ \cline{2-7}

& MOON
& 24.0316 / 0.8582 & 24.1474 / 0.8491 & 25.1535 / 0.9038 & \textbf{34.6905 / 0.9985}
& 27.0058 / 0.9024 \\ \cline{2-7}

& FedDoM
& \textbf{25.6286 / 0.8962} & \textbf{25.7792 / 0.8876} & \textbf{26.6404 / 0.9314} & 31.9686 / 0.9962
& \textbf{27.5042 / 0.9279} \\ \hline

\multirow{4}{*}{7} 
& FedAvg
& 23.4463 / 0.8509 & 23.4762 / 0.8383 & 24.4676 / 0.8914 & 34.7323 / 0.9986
& 26.5306 / 0.8948 \\ \cline{2-7}

& FedProx
& 24.8811 / 0.8929 & 24.9595 / 0.8837 & 25.6997 / 0.9240 & 35.4577 / 0.9989
& 27.7495 / 0.9249 \\ \cline{2-7}

& MOON
& 25.3384 / 0.9021 & 25.4905 / 0.8921 & 26.2142 / 0.9313 & \textbf{36.0850 / 0.9990}
& 28.2820 / 0.9312 \\ \cline{2-7}

& FedDoM
& \textbf{26.8932 / 0.9295} & \textbf{27.0741 / 0.9220} & \textbf{27.8017 / 0.9530} & 33.0390 / 0.9975
& \textbf{28.7020 / 0.9505} \\ \hline

\multirow{4}{*}{10} 
& FedAvg
& 24.1878 / 0.8866 & 24.1791 / 0.8730 & 24.9474 / 0.9102 & 35.5192 / 0.9989
& 27.2084 / 0.9172 \\ \cline{2-7}

& FedProx
& 25.6403 / 0.9202 & 25.6984 / 0.9104 & 26.2041 / 0.9402 & 36.3133 / 0.9992
& 28.4640 / 0.9425 \\ \cline{2-7}

& MOON
& 26.1896 / 0.9264 & 26.3605 / 0.9164 & 26.8509 / 0.9469 & \textbf{36.9687 / 0.9993}
& 29.0924 / 0.9472 \\ \cline{2-7}

& FedDoM
& \textbf{27.7384 / 0.9478} &\textbf{ 27.9441 / 0.9405} & \textbf{28.4965 / 0.9648} & 33.6722 / 0.9981
& \textbf{29.4628 / 0.9628} \\ \hline

\multirow{4}{*}{13} 
& FedAvg
& 24.5417 / 0.9042 & 24.5065 / 0.8898 & 25.1650 / 0.9198 & 35.9707 / 0.9991
& 27.5460 / 0.9282 \\ \cline{2-7}

& FedProx
& 26.0381 / 0.9332 & 26.0889 / 0.9229 & 26.4653 / 0.9481 & 36.8105 / 0.9993
& 28.8507 / 0.9509 \\ \cline{2-7}

& MOON
& 26.6656 / 0.9378 & 26.8409 / 0.9279 & 27.2035 / 0.9547 & \textbf{37.4750 / 0.9994}
& 29.5463 / 0.9549 \\ \cline{2-7}

& FedDoM
& \textbf{28.2225 / 0.9568} & \textbf{28.4470 / 0.9495} & \textbf{28.8829 / 0.9705} & 34.0293 / 0.9983
& \textbf{29.8954 / 0.9688} \\ \hline

\end{tabular}}
\label{UnbalancedResult}
\end{table*}

\subsection{Performance of the Trained Model under Rayleigh Channel}
Fig.~\ref{rayleigh} illustrates the PSNR and MS-SSIM performance across different SNR values for the semantic communication trained by four FL training frameworks over the Photo, Art, Cartoon, and Sketch domains. The fading coefficient elements follow the $\mathcal{CN}(0,\sigma^{2}_{k})$ distribution for the Rayleigh channel. Overall, the proposed framework consistently achieves the highest reconstruction quality in most domains and SNR conditions, confirming the effectiveness of the global representation alignment and domain-aware aggregation mechanisms. Compared with the AWGN channel, the overall performance is slightly lower due to signal fading in the wireless channel; however, it still maintains high performance under low SNR conditions and further improves reconstruction quality as the noise level decreases. Specifically, in the \textbf{Cartoon} and \textbf{Photo domains}, the improvement over the second best is more evident, $0.4278$ and $0.5032$ at $1$ dB; $0.5899$ and $0.5049$ for PSNR at $13$ dB, which demonstrate the robustness to channel variations through feature generalization and alignment. In the \textbf{Art domain}, all methods perform well due to its simpler structure, yet our approach still maintains a clear advantage. Finally, for the \textbf{Sketch domain}, the FedAvg framework secures the highest performance due to two main reasons: unconstrained local training and the dominance of the Sketch domain in the global aggregation process, resulting from its large amount of data samples. The proposed FedDoM obtains lower performance in this domain but secures outstanding performance in the remaining domains and in the overall average across domains, indicating a trade-off between domain-specific performance and global generalization.

\subsection{Performance under the heterogeneity of domain data}\label{dominancedomaindiscussion}
Up to this point, we have conducted experiments for the different FL training frameworks with the $30\%$ data sample from all domains. However, this approach is relatively naive when considering the difficulty of data generation for each domain. For example, for an art painting, it can take months or even years to produce one sample, while the sketch image is relatively easy and needs much less effort. Therefore, to further emphasize this property of real-world environment and demonstrate the effectiveness of our proposed framework, we sample $20\%$ of data for three domains: Photo, Art painting, and Cartoon, while $50\%$ for the Sketch domain. The number of data samples is provided in detail in Table~\ref{unbalanced}.

In Table~\ref{UnbalancedResult}, we present the simulation results when the heterogeneity among domains becomes more extreme. As shown, the performance of MOON, FedProx, and FedAvg on low-sample domains-Photo, Art Painting, and Cartoon-degrades significantly compared to the earlier training setup in Table~\ref{PerformanceAWGN}. In contrast, the proposed framework is able to preserve semantic information across domains during both local training and global aggregation, leading to strong performance even for low-sample domains. Specifically, FedDoM outperforms the second-best framework (MOON) by $1.7924$, $1.8220$, and $1.3673$ in PSNR, and by $0.0562$, $0.0547$, and $0.0352$ in MS-SSIM under the SNR of $1$ dB. These performance gaps further widen as the channel noise decreases, demonstrating the ability of the semantic communication model and training framework to adapt effectively to improved channel conditions.

In general, our framework achieves higher performance on the three low-sample domains; however, this comes at the expense of reduced performance on the high-sample domain (Sketch) compared to the other FL frameworks. This behavior can be attributed to the dominance of the Sketch domain during the aggregation process, where its contribution becomes disproportionately large, as reflected in Equation (\ref{FedAvg}) for the FedAvg, FedProx, and MOON frameworks. In contrast, our approach establishes a more balanced and domain-aware training process that performs well across all domains and achieves the highest values for both evaluation metrics when averaging over all domains, as shown in the last columns of Table~\ref{UnbalancedResult}.

\vspace{-3mm}
\section{Conclusion}\label{Conclusion}

In this paper, we have introduced a novel federated learning framework to address the domain shift problem when client data originates from various sources in training a deep learning-based joint source-channel coding model. The difference in domains of client data can lead to significant variation in learning parameters among client models, which degrades the performance of the global model. Therefore, we have constructed a global feature representation from the local features at the server, which effectively provides a soft boundary for the local training at the clients. We have leveraged this global representation to align the local features, which prevents parameter updates from drifting too far from the generalization target. We have optimized the local model by combining the generalization loss with the local reconstruction loss, thereby facilitating both local domain learning and cross-domain learning. On the other hand, we have considered the heterogeneity in terms of data samples from different domains and proposed a domain-aware aggregation approach to prevent the dominance of easily generated data domains in the global aggregation process. Finally, the experimental results verify that our framework consistently outperforms benchmark methods in three out of four domains, while still achieving strong performance in the remaining one, highlighting its robustness and generalization capability across diverse data distributions.
\vspace{-5mm}

\bibliographystyle{IEEEtran}
\bibliography{mybib}

@ARTICLE{10856890,
  author={Jia, Ninghui and Qu, Zhihao and Ye, Baoliu and Wang, Yanyan and Hu, Shihong and Guo, Song},
  journal={IEEE Commun. Surveys Tuts.}, 
  title={A Comprehensive Survey on Communication-Efficient Federated Learning in Mobile Edge Environments}, 
  year={Early Access, 2025},
  doi={10.1109/COMST.2025.3535957}}

@ARTICLE{11193869,
  author={Zhou, Conghao and Gao, Jie and Hu, Shisheng and Cheng, Nan and Zhuang, Weihua and Shen, Xuemin},
  journal={IEEE Trans. Mob. Comput.}, 
  title={User-Centric Communication Service Provision for Edge-Assisted Mobile Augmented Reality}, 
  year={Early Access, 2025},
  doi={10.1109/TMC.2025.3618147}}

@ARTICLE{11192484,
  author={Nguyen, Loc X. and others},
  journal={IEEE Commun. Surveys Tuts.}, 
  title={A Contemporary Survey on Semantic Communications: Theory of Mind, Generative AI, and Deep Joint Source-Channel Coding}, 
  year={Early Access, 2025},
  doi={10.1109/COMST.2025.3616973}}

@ARTICLE{10930485,
  author={Pivoto, Diego Gabriel Soares and de Figueiredo, Felipe A. P. and others},
  journal={IEEE Commun. Surveys Tuts.}, 
  title={A Comprehensive Survey of Machine Learning Applied to Resource Allocation in Wireless Communications}, 
  year={Mar. 2025, Early Access},
  doi={10.1109/COMST.2025.3552370}}

@ARTICLE{10944429,
  author={Guo, Jie and Yin, Hang and Song, Bin and Chi, Yuhao and Zhang, Zhaoyang and others},
  journal={IEEE Trans. Wirel. Commun.}, 
  title={Multi-Scale Semantic Communication for Object Detection: Single and Cross-Domain Scenarios}, 
  year={Mar. 2025},
  volume={24},
  number={7},
  pages={6195-6210},
  doi={10.1109/TWC.2025.3552255}}

@ARTICLE{10552845,
  author={Hou, Ricky Yuen-Tan and Liu, Gang and Fong, Jefferson and Zhang, Hui and Jeong, Seon-Phil},
  journal={IEEE Access}, 
  title={A Model for Quantifying the Degree of Understanding in Cross-Domain M2M Semantic Communications}, 
  year={Jun. 2024},
  volume={12},
  pages={84515-84536},
  doi={10.1109/ACCESS.2024.3412929}}

@ARTICLE{10041216,
  author={Hu, Chung-Hsuan and Chen, Zheng and Larsson, Erik G.},
  journal={IEEE J. Sel. Areas Commun.}, 
  title={Scheduling and Aggregation Design for Asynchronous Federated Learning Over Wireless Networks}, 
  year={Feb. 2023},
  volume={41},
  number={4},
  pages={874-886},
  doi={10.1109/JSAC.2023.3242719}}

@ARTICLE{9562522,
  author={Lin, Frank Po-Chen and Hosseinalipour, Seyyedali and Azam, Sheikh Shams and Brinton, Christopher G. and Michelusi, Nicolò},
  journal={IEEE J. Sel. Areas Commun.}, 
  title={Semi-Decentralized Federated Learning With Cooperative D2D Local Model Aggregations}, 
  year={Oct. 2021},
  volume={39},
  number={12},
  pages={3851-3869},
  doi={10.1109/JSAC.2021.3118344}}

@ARTICLE{9724245,
  author={Zhang, Haobo and Zhang, Hongliang and Di, Boya and Renzo, Marco Di and Han, Zhu and Poor, H. Vincent and Song, Lingyang},
  journal={IEEE J. Sel. Areas Commun.}, 
  title={Holographic Integrated Sensing and Communication}, 
  year={Mar. 2022},
  volume={40},
  number={7},
  pages={2114-2130},
  doi={10.1109/JSAC.2022.3155548}}

@book{shannon1998mathematical,
  title={The mathematical theory of communication},
  author={Shannon, Claude E and Weaver, Warren},
  year={1998},
  publisher={University of Illinois press}
}

@ARTICLE{10531097,
  author={Nguyen, Loc X. and Le, Huy Q. and Tun, Ye Lin and Aung, Pyae Sone and Tun, Yan Kyaw and Han, Zhu and Hong, Choong Seon},
  journal={IEEE Trans. Veh. Technol.}, 
  title={An Efficient Federated Learning Framework for Training Semantic Communication Systems}, 
  year={May 2024},
  volume={73},
  number={10},
  pages={15872-15877},
  doi={10.1109/TVT.2024.3401140}}

@ARTICLE{10288204,
  author={Picano, Benedetta and Fantacci, Romano},
  journal={IEEE Internet Things J.}, 
  title={A Semantic-Oriented Federated Learning for Hybrid Ground–Aqua Computing Systems}, 
  year={Mar. 2024},
  volume={11},
  number={6},
  pages={10095-10103},
  doi={10.1109/JIOT.2023.3325289}}

@INPROCEEDINGS{9685654,
  author={Tong, Haonan and Yang, Zhaohui and Wang, Sihua and Hu, Ye and Saad, Walid and Yin, Changchuan},
  booktitle={Proc. IEEE Glob. Commun. Conf.}, 
  title={Federated Learning based Audio Semantic Communication over Wireless Networks}, 
  year={Marid, Spain, Dec. 2021},
  doi={10.1109/GLOBECOM46510.2021.9685654}}

@ARTICLE{10559618,
  author={Wang, Yining and Ni, Wanli and Yi, Wenqiang and Xu, Xiaodong and Zhang, Ping and Nallanathan, Arumugam},
  journal={IEEE Commun. Lett.}, 
  title={Federated Contrastive Learning for Personalized Semantic Communication}, 
  year={Jun. 2024},
  volume={28},
  number={8},
  pages={1875-1879},
  doi={10.1109/LCOMM.2024.3415431}}

@inproceedings{li2021model,
  title={Model-contrastive federated learning},
  author={Li, Qinbin and He, Bingsheng and Song, Dawn},
  booktitle={Proc. IEEE Comput. Soc. Conf. Comput. Vis. Pattern Recognit. (CVPR)},
  year={Virtual Conference, Jun. 2021}
}

@article{wu2025personalized,
  title={Personalized federated learning for semantic communication with collaborative fine-tuning},
  author={Wu, Maochuan and Li, Juan and Xu, Jing and Chen, Bing and Zhu, Kun},
  journal={Digit. Commun. Netw.},
  year={Early Access, Aug. 2025},
  publisher={Elsevier}
}

@inproceedings{mcmahan2017communication,
  title={Communication-efficient learning of deep networks from decentralized data},
  author={McMahan, Brendan and Moore, Eider and Ramage, Daniel and others},
  booktitle={Proc. Int. Conf. Artif. Intell. Statist. (AISTATS)},
  year={Fort Lauderdale, FL, Apr. 2017},
}

@article{DJSCCnofeedback,
  title={Deep joint source-channel coding for wireless image transmission},
  author={Bourtsoulatze, Eirina and Kurka, David Burth and G{\"u}nd{\"u}z, Deniz},
  journal={IEEE Trans. Cogn. Commun. Netw.},
  volume={5},
  number={3},
  pages={567--579},
  year={Sep. 2019},
  publisher={IEEE}
}

@ARTICLE{DJSCCwithfeedback,
  author={Kurka, David Burth and Gündüz, Deniz},
  journal={IEEE J. Sel. Areas Inf. Theory}, 
  title={DeepJSCC-f: Deep Joint Source-Channel Coding of Images With Feedback}, 
  year={May 2020},
  volume={1},
  number={1},
  pages={178-193},
  doi={10.1109/JSAIT.2020.2987203}}

@INPROCEEDINGS{DJSCCSwin,
  author={Yang, Ke and Wang, Sixian and others},
  booktitle={Proc. IEEE Int. Conf. Acoust. Speech Signal Process. (ICASSP)}, 
  title={WITT: A Wireless Image Transmission Transformer for Semantic Communications}, 
  year={Rhodes Island, Greece, Jun. 2023},
  doi={10.1109/ICASSP49357.2023.10094735}}

@article{DJSCCAttention,
  title={Wireless image transmission using deep source channel coding with attention modules},
  author={Xu, Jialong and Ai, Bo and Chen, Wei and Yang, Ang and others},
  journal={IEEE Trans. Circuits Syst. Video Technol.},
  volume={32},
  number={4},
  pages={2315--2328},
  year={May 2021},
  publisher={IEEE}
}

@ARTICLE{DJSCCDigital,
  author={Park, Joohyuk and Oh, Yongjeong and Kim, Seonjung and Jeon, Yo-Seb},
  journal={IEEE Trans. Cogn. Commun. Netw.}, 
  title={Joint Source-Channel Coding for Channel-Adaptive Digital Semantic Communications}, 
  year={Feb. 2025},
  volume={11},
  number={1},
  pages={75-89},
  doi={10.1109/TCCN.2024.3422496}}

@inproceedings{liu2021swin,
  title={Swin transformer: Hierarchical vision transformer using shifted windows},
  author={Liu, Ze and Lin, Yutong and Cao, Yue and Hu, Han and Wei, Yixuan and Zhang, Zheng and Lin, Stephen and et al.},
  booktitle={Proc. IEEE Int. Conf. Comput. Vis.},
  year={Montreal, Canada, Oct. 2021}
}

@article{FLSemantic,
  title={Federated Learning for Semantic Communication Based on CNNs and Transformer},
  author={Li, Shufeng and Cai, Yujun and Deng, Zhaokai and Ba, Xinran and Zheng, Qinghe and Zhang, Xinruo and Su, Baoxin},
  journal={Int. J. Intell. Syst.},
number = {1},
pages = {1-22},
  year={Aug. 2025},
  publisher={Wiley Online Library}
}

@ARTICLE{FLforGAIassistedSemCom,
  author={Peng, Yubo and Jiang, Feibo and Dong, Li and Wang, Kezhi and Yang, Kun},
  journal={IEEE Trans. Cogn. Commun. Netw.}, 
  title={Personalized Federated Learning for GAI-Assisted Semantic Communications}, 
  year={Early Access, Jul. 2025},
  doi={10.1109/TCCN.2025.3586904}}

@INPROCEEDINGS{FLfortaskoriented,
  author={Sun, Haofeng and Tian, Hui and Ni, Wanli and Zheng, Jingheng},
  booktitle={Proc. IEEE Conf. Comput. Commun. Workshops  (INFOCOM WKSHPS)}, 
  title={Federated Learning-Based Cooperative Model Training for Task-Oriented Semantic Communication}, 
  year={Vancouver, Canada, May 2024},
  doi={10.1109/INFOCOMWKSHPS61880.2024.10620817}}

@article{huh2025feature,
  title={Feature Reconstruction Aided Federated Learning for Image Semantic Communication},
  author={Huh, Yoon and Kim, Bumjun and Choi, Wan},
  journal={Proc. IEEE Glob. Commun. Conf.},
  year={Taipei, Taiwan, China, Dec. 2025}
}

@ARTICLE{10937229,
  author={Sun, Haofeng and Ni, Wanli and Tian, Hui and Zheng, Jingheng and Nie, Gaofeng and Niyato, Dusit},
  journal={IEEE Internet Things J.}, 
  title={A Hybrid Federated Learning Framework for Task-Oriented Semantic Communication}, 
  year={Mar. 2025},
  volume={12},
  number={13},
  pages={23444-23461},
  doi={10.1109/JIOT.2025.3553504}}

@ARTICLE{11096603,
  author={Huang, Ruochen and Qiu, Chen and Chen, Mingkai and Zhang, Changwei and Zhu, Hongbo},
  journal={IEEE Internet Things J.}, 
  title={Federated-Learning-Enabled Cross-Modal Semantic Communication for 6G}, 
  year={Jul. 2025},
  volume={12},
  number={20},
  pages={41608-41624},
  doi={10.1109/JIOT.2025.3590596}}

@article{zhou2025fedsem,
  title={FedSem: A Resource Allocation Scheme for Federated Learning Assisted Semantic Communication},
  author={Zhou, Xinyu and Li, Yang and Zhao, Jun},
  journal={ArXiv:2503.06058},
  year={Mar. 2025}
}

@ARTICLE{10960416,
  author={Won, Dongwook and Do, Quang Tuan and Win, Thwe Thwe and Lee, Donghyun and Oh, Junsuk and Cho, Sungrae},
  journal={IEEE J. Sel. Areas Commun.}, 
  title={Multidomain Adaptive Semantic Communications}, 
  year={Apr. 2025},
  volume={43},
  number={7},
  pages={2506-2517},
  doi={10.1109/JSAC.2025.3559127}}

@inproceedings{li2017deeper,
  title={Deeper, broader and artier domain generalization},
  author={Li, Da and Yang, Yongxin and Song, Yi-Zhe and Hospedales, Timothy M},
  booktitle={Proc. of the IEEE Int. Conf. Comput. Vis.},
  year={Venice, Italy, Oct. 2017}
}

@article{li2020federated,
  title={Federated optimization in heterogeneous networks},
  author={Li, Tian and Sahu, Anit Kumar and Zaheer, Manzil and Sanjabi, Maziar and Talwalkar, Ameet and Smith, Virginia},
  journal={in Proc. Mach. Learn. Syst},
  volume={2},
  pages={429--450},
  year={Austin, TX, Mar. 2020}
}

@article{qiao2025towards,
  title={Towards artificial general or personalized intelligence? A survey on foundation models for personalized federated intelligence},
  author={Qiao, Yu and Le, Huy Q and Raha, Avi Deb and Tran, Phuong-Nam and others},
  journal={ArXiv:2505.06907},
  year={2025}
}

@article{qiao2025deepseek,
author = {Qiao, Yu and Tran, Phuong-Nam and Yoon, Ji and others},
title = {DeepSeek-Inspired Exploration of RL-Based LLMs and Synergy with Wireless Networks: A Survey},
year = {Early Access, Nov. 2025},
publisher = {Association for Computing Machinery},
address = {New York, NY, USA},
issn = {0360-0300},
doi = {10.1145/3776745},
journal = {ACM Comput. Surv.}
}

@inproceedings{pillutla2022federated,
  title={Federated learning with partial model personalization},
  author={Pillutla, Krishna and Malik, Kshitiz and Mohamed, Abdel-Rahman and Rabbat, Mike and Sanjabi, Maziar and Xiao, Lin},
  booktitle={Proc. Int. Conf. Mach. Learn.},
  year={Baltimore, Maryland, Jul. 2022},
  organization={PMLR}
}

@INPROCEEDINGS{10203389,
  author={Huang, Wenke and Ye, Mang and Shi, Zekun and others},
  booktitle={Proc. IEEE/CVF Comput. Soc. Conf. Comput. Vis. Pattern Recognit. (CVPR)}, 
  title={Rethinking Federated Learning with Domain Shift: A Prototype View}, 
  year={BC, Canada, Jun. 2023},
  doi={10.1109/CVPR52729.2023.01565}}

@inproceedings{wang2025federated,
  title={Federated Learning with Domain Shift Eraser},
  author={Wang, Zheng and Wang, Zihui and Fan, Xiaoliang and Wang, Cheng},
  booktitle={Proc. of the Comput. Vis. Pattern Recognit. (CVPR)},
  year={Nashville, Tennessee, Jun. 2025}
}

@ARTICLE{10729265,
  author={Tan, Weixing and Wang, Yusen and Liu, Lei and others},
  journal={IEEE Internet Things J.}, 
  title={Adaptive Federated Deep Learning-Based Semantic Communication in the Social Internet of Things}, 
  year={Sep. 2025},
  volume={12},
  number={17},
  pages={34921-34930},
  doi={10.1109/JIOT.2024.3484230}}

@inproceedings{chen2024fair,
  title={Fair federated learning under domain skew with local consistency and domain diversity},
  author={Chen, Yuhang and Huang, Wenke and Ye, Mang},
  booktitle={Proc. of the IEEE/CVF Conf. Comput. Vis. Pattern Recognit.},
  year={WA, USA, Jun. 2024}
}

@inproceedings{mohri2019agnostic,
  title={Agnostic federated learning},
  author={Mohri, Mehryar and Sivek, Gary and Suresh, Ananda Theertha},
  booktitle={Proc. of the 36th Int. Conf. Mach. Learn. (ICML)},
  year={Long Beach, California, Jun. 2019},
  organization={PMLR}
}

@article{wang2024taming,
  title={Taming cross-domain representation variance in federated prototype learning with heterogeneous data domains},
  author={Wang, Lei and Bian, Jieming and Zhang, Letian and Chen, Chen and Xu, Jie},
  journal={in Proc. Adv. Neural Inf. Process. Syst. (NeurIPS)},
  volume={37},
  year={Vancouver, Canada, Dec. 2024}
}

@inproceedings{tan2022fedproto,
  title={Fedproto: Federated prototype learning across heterogeneous clients},
  author={Tan, Yue and Long, Guodong and Liu, Lu and Zhou, Tianyi and Lu, Qinghua and others},
  booktitle={Proc. AAAI Conf. Artif. Intell.},
  volume={36},
  number={8},
  year={Virtual Conference, Feb. 2022}
}

@ARTICLE{10929033,
  author={Saad, Walid and Hashash, Omar and Thomas, Christo Kurisummoottil and Chaccour, Christina and Debbah, Mérouane and Mandayam, Narayan and Han, Zhu},
  journal={Proceedings of the IEEE}, 
  title={Artificial General Intelligence (AGI)-Native Wireless Systems: A Journey Beyond 6G}, 
  year={Mar. 2025},
  doi={10.1109/JPROC.2025.3526887}}

@ARTICLE{8755300,
  author={Chen, Mingzhe and Challita, Ursula and Saad, Walid and others},
  journal={IEEE Commun. Surveys Tuts.}, 
  title={Artificial Neural Networks-Based Machine Learning for Wireless Networks: A Tutorial}, 
  year={Fourthquarter 2019},
  volume={21},
  number={4},
  pages={3039-3071},
  doi={10.1109/COMST.2019.2926625}}

@ARTICLE{10554663,
  author={Chaccour, Christina and Saad, Walid and Debbah, Mérouane and others},
  journal={IEEE Commun. Surveys Tuts.}, 
  title={Less Data, More Knowledge: Building Next-Generation Semantic Communication Networks}, 
  year={Firstquarter 2025},
  volume={27},
  number={1},
  pages={37-76},
  doi={10.1109/COMST.2024.3412852}}

\vfill

\end{document}